\documentclass[traditabstract]{aa} 	% for the abstract without structuration 
\usepackage{graphicx}
\usepackage{txfonts}
\usepackage{natbib}
\bibpunct{(}{)}{;}{a}{}{,} % to follow the A&A style
\begin{document}

   \title{The stellar correlation function from SDSS}

   \subtitle{A statistical search for wide binary stars}

   \author{M. Longhitano\and B. Binggeli}

   \institute{Department of Physics, Basel University,
              Klingelbergstrasse 82, CH-4056 Basel\\
              \email{marco.longhitano@unibas.ch}}

   \date{Received 12 August 2009 / Accepted 1 October 2009}

\abstract
{We study the statistical properties of the wide binary population in the Galaxy field with projected separations larger than 200 AU by constructing the stellar \emph{angular} two-point correlation function (2PCF) from a homogeneous sample of nearly 670\,000 main sequence stars. The selected stars lie within a rectangular region around the Northern Galactic Pole and have apparent $r$-band magnitudes between 15 and 20.5 mag and spectral classes later than G5 ($g-r>0.5$ mag). The data were taken from the Sixth Data Release of the Sloan Digital Sky Survey. We model the 2PCF by means of the Wasserman-Weinberg technique including several assumptions on the distribution of the binaries' orbital parameters, luminosity function, and density distribution in the Galaxy. In particular, we assume that the semi-major axis distribution is described by a single powerlaw.
The free model parameters -- the local wide binary number density $n_{\rm WB}$ and the power-law index $\lambda$ of the semi-major axis distribution -- are inferred simultaneously by least-square fitting.
We find the separation distribution to follow \"Opik's law ($\lambda=1$) up to the Galactic tidal limit, without any break and 
a local density of 5 wide binaries per 1\,000 pc$^3$ with both components having spectral type later than G5. This implies that about 10\% of all stars in the solar neighbourhood are members of such a late-type wide binary system.
With a relative statistical (2$\sigma$) error of about 10\%, our findings are in general agreement with previous studies of wide binaries.
   The data suggest that about 800 very wide pairs with projected separations larger than 0.1 pc exist in our sample, whereas none are found beyond 0.8 pc.
   Modern large-scale surveys make the 2PCF method a viable tool for studying wide binary stars and a true complement to common proper motion studies. The method is, however, seriously limited by the noise from optical pairs and the (over)simplifying assumptions made to model the selection effects and to interpret the measured clustering signal. 
}
 

   \maketitle
%________________________________________________________________

\section{Introduction}
Binary stars have traditionally been central for astronomy, especially {\em close} binary systems (with typical orbital periods of days to years), because they are genuine laboratories of stellar evolution and its exotic remnants, making them a cornerstone for determining masses, distances, and many other fundamental astrophysical parameters. On the other hand, {\em wide} binary systems (orbital periods of many thousands to millions of years) are of particular interest, too.

Because they are only weakly bound by gravitation, wide binaries are prone to tidal disruption by passing massive objects, such as massive stars, molecular clouds, MACHOs (massive compact halo objects), or dark matter (DM) substructure. The shape of their separation distribution, in particular towards the most extreme, widest separations, should therefore allow constraints on the mass and frequency of the disruptive perturbers \citep{1982ApJ...254..214R}, as well as to estimate the age of a population \citep{2004RMxAC..21...49P}. Wide binary-based MACHO constraints have been placed (but subsequently criticised) by several authors \citep{1985ApJ...290...15B,1987ApJ...312..367W,1990bdm..proc..117W,1991ApJ...382..149W,2004ApJ...601..311Y,2009MNRAS.396L..11Q}.

In a different context, \citet{2008MNRAS.387.1727H} propose using the tightening of wide binaries in dwarf spheroidal galaxies through dynamical friction as a test for DM. 
Moreover, binaries with separations over 0.1 pc, which are known to exist in the Galaxy \citep[e.g.][also this paper]{2007AJ....133..889L}, are in the `weak-accelaration' regime where, in principle, one could test for possible deviations from Newtonian gravity, such as MOND \citep{1987IAUS..117..319M,1990AJ....100.1968C}.

Wide binaries are not only probes of dynamical evolution and Galactic structure, but they also provide important clues to star formation \citep{2001IAUS..200...93L}. The exact outcome of the binary population in a star-formation event is an exceedingly complex and still unsolved problem \citep[][and references therein]{2007prpl.conf..133G}. The formation of extremely wide binaries is particularly difficult to understand (\citeauthor{2007IAUS..240..405A}~\citeyear{2007IAUS..240..405A};~\citeauthor{2009MNRAS.tmp..873P}~\citeyear{2009MNRAS.tmp..873P};~ \citeauthor{2010arXiv1001.3969K}~\citeyear{2010arXiv1001.3969K}). 
A good knowledge of the wide binary frequency and separation distribution is primary for a whole host of problems in astrophysics \citep[e.g.][]{2007IAUS..240..316C}.

Unfortunately, the long orbital periods of wide binary systems make their identification very difficult in the first place. 
There are basically two different methods for detecting a wide binary system: (1) by the number excess of neighbours around a given star with respect to a random distribution, or (2) by the common proper motion (CPM) of two well-separated stars. Although there is no way to observe the orbital motion of a wide binary pair, the CPM is nevertheless a reliable indicator of a physical relationship between two individual stars \citep[e.g.][]{2007AJ....133..889L}. In the present study we use the {\em angular two-point correlation function} (2PCF) to measure the excess of neighbours compared to a random distribution. This method has the advantage that larger samples of more distant stars can be used, as only the stars' positions are required. It only allows, however, statistical statements on the pairing and is limited to relatively small angular separations, because the noise due to randomly associated pairs increases rapidly (linearly) with angular separation. 

To date, our knowledge on wide binaries essentially had rested on studies using the CPM method. There are two excellent recent CPM studies of wide binaries by \citet{2004ApJ...601..289C} and \citet{2007AJ....133..889L}. Based on updated proper motion (Luyten and Hipparcos) catalogues of stars, 917 and 521 wide binary systems, respectively, have been identified over the entire Northern sky and with a typical median distance of about 100 pc. \citeauthor{2007AJ....133..889L} found the separation distribution of the pairs to follow \citeauthor{1924PTarO..25f...1O}'s (\citeyear{1924PTarO..25f...1O}) law, i.e. frequency being proportional to the inverse of separation, out to separations of around 3\,500 AU. Beyond this characteristic scale, however, the separation distribution seems to be falling by a steeper power law, without an obvious cut-off. \citeauthor{2007AJ....133..889L} find wide binaries out to separations of almost 100\,000 AU, or 0.4 pc! They quote a number of 9.5\% for nearby ($D<100$ pc) Hipparcos stars belonging to a wide binary system with a separation greater than 1\,000 AU, again demonstrating the ubiquity of the (wide) binary phenomenon. \citeauthor{2004ApJ...601..289C}, in addition, achieve a distinction between wide binaries belonging to the Galactic disc and those belonging to the Galactic halo. No significant difference in the separation distribution has been found, suggesting that the disc and halo binaries were formed under similar conditions, despite the very different metallicities and ages. 

The angular 2PCF (also called ``covariance function'') is one of the most useful tools for studying the clustering properties of {\em galaxies}, and it has been widely used to probe the large-scale structure of the Universe~\citep[e.g.][for a more recent study of the angular clustering of galaxies see~\citeauthor{2002ApJ...579...48S}~\citeyear{2002ApJ...579...48S} and references therein.]{1980lssu.book.....P}. The 2PCF can also be, and has been, used to measure the clustering properties of stars. While on large scales the stars are clearly randomly distributed in the Galaxy, as expected in a collisionless system, correlations up to the 10 pc scale (or even beyond) can be found in moving groups and halo streamers~\citep{1989ApJ...340L..57D}, star-forming regions~\citep[e.g.][]{1998AJ....115.1524G}, and open star clusters~\citep{1998MNRAS.301..289L}. On very small scales (sub-pc, observationally: sub-arcmin), there is a strong signal due to (visual) binary stars.

The 2PCF method of probing wide binary stars in the Galactic field was pioneered by \citet{1981ApJ...246..122B}, who studied the distribution of stars down to a limiting magnitude V = 16 in a 10 square degree field at the NGP, and found very significant clustering at a separation of 0.1 pc. Of the 19 binary candidates, 6 turned out to be real \citep{1984ApJ...281L..41L}. The theoretical implications of these observations for the frequency and separation distribution of the binaries, and a general method for modelling them, was worked out by \citet{1987ApJ...312..390W}. To date, there are only few follow-up studies of the stellar 2PCF \citep{1989MNRAS.237..311S,1988ApJ...335L..47G,1991PhDT.........8G,1995ApJ...441..200G}. 
Given this surprising lack of further work on the stellar 2PCF to study wide binaries, and in view of the enormous progress in deep photometric sky surveys in the past 20 years that should render the 2PCF method -- as a true complement to the CPM method -- much more rewarding now, we have started a project to use the huge stellar database of the Sloan Digital Sky Survey \citep[SDSS][\texttt{www.sdss.org}]{2000AJ....120.1579Y} for a detailed stellar correlation analysis to very faint magnitudes. 

An independent study of wide binaries in the SDSS database by \citet{2008ApJ...689.1244S} takes a different approach, based on the ``Milky Way Tomography'' of \citet{2008ApJ...673..864J}, where approximate distances are ascribed to all stars by adopting a photometric parallax relation. Candidate binaries are selected by the requirement that the difference between two potential components in apparent magnitude is within a certain error equal to the difference in absolute magnitude. 

Using distance information has the fundamental advantage of filtering out the disturbing noise from chance projections and of avoiding the need for a complex, \citeauthor{1987ApJ...312..390W}-type modelling to calculate integrated, sky-projected quantities. Although we plan to include distance information in future work, we show here that the {\em angular} correlation analysis is in principle still a viable approach. Our main results agree with those of the CPM studies. We also show, however, what the limitations of the method are.
         
The paper is organised as follows. In Sect.~\ref{sample} we describe the SDSS data input and define the cleansed sample used for the analysis. Section \ref{scf} is devoted to the angular correlation function apparatus, followed by an extensive description of the modelling of the correlation function drawing on a modified \citeauthor{1987ApJ...312..390W} technique in Sect.~\ref{model}. Our results for the total sample and for a number of subsamples are presented in Sect.~\ref{results} and are critically discussed and compared to previous work in Sect.~\ref{discuss}. Concluding remarks are given in Sect.~\ref{conclude}. 
%__________________________________________________________________

\section{Data}\label{sample}
% Fig. 1, starDist
   \begin{figure*}
   \resizebox{\hsize}{!}{\includegraphics{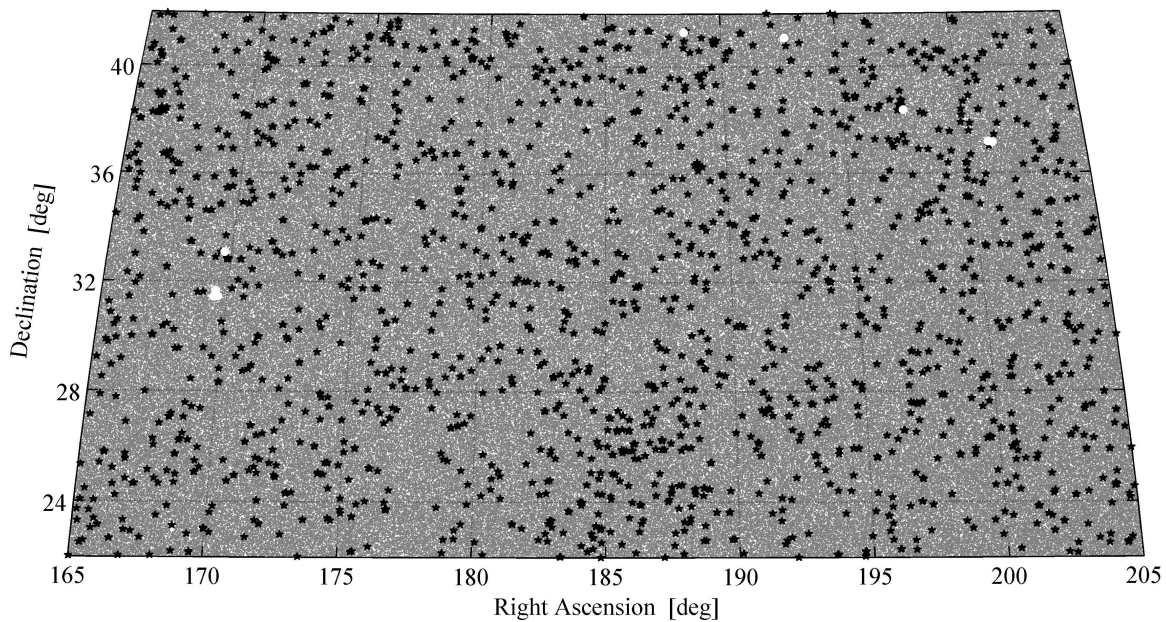}}
      \caption{Distribution of stars (grey points filling the background) in our total sample. Black asterisks show the positions of bright star masks, white circles those of hole masks. The size of the symbols are not to scale.}
         \label{starDist}
   \end{figure*}   
For our analysis we selected a homogeneous sample of stars from the Sixth SDSS Data
Release~\citep[DR6;][]{2008ApJS..175..297A}.
We took a rectangular area around the Northern Galactic Pole (NGP) covering approximately
$\Omega_{\rm tot}\simeq675$ square degrees (dec: $22^{\circ} - 42^{\circ}$, RA: $165^{\circ} - 205^{\circ}$; see Fig.~\ref{starDist}). 
It contains 966\,656 \emph{primary}\footnote{Due to overlaps in the imaging, many objects are observed more than once. The best of those observations is called the \emph{primary object}, all the others are called \emph{secondary objects}.} point-like objects
-- including quasars, asteroids, and possibly some misidentified galaxies -- selected from the \emph{Star} view\footnote{See 
\texttt{cas.sdss.org/dr6/en/help/browser/browser.asp}}, having an apparent PSF
magnitude\footnote{See 
\texttt{sdss.org/dr6/algorithms/photometry.html\#mag\_psf}} in the $r$-band between 15 
and 20.5 mag.
Following the SDSS recommendations\footnote{See 
\texttt{sdss.org/dr6/products/catalogs/flags.html}},
we required the stars to have \emph{clean photometry}\footnote{See 
\texttt{cas.sdss.org/dr6/en/help/docs/realquery.asp\#flags}} in the $g$, $r$, and $i$ band.
Although the effect of interstellar dust on the measurements is waek in the direction of the NGP, we corrected the data using the~\citet{1998ApJ...500..525S} maps, which can be easily done, as the \texttt{extinction} at the position of each object is stored in the SDSS database. 

The magnitude limits were chosen to be within the saturation limit of the SDSS CCD
cameras~\citep[$r\sim 14$ mag;][]{1998AJ....116.3040G}\footnote{See also 
\texttt{astro.princeton.edu/PBOOK/camera/camera.htm}}
and the limit where the star-galaxy separation becomes
unreliable~\citep[$r\sim 21.5$ mag;][]{2001ASPC..238..269L}\footnote{See also 
\texttt{sdss.org/dr6/products/general/stargalsep.html}}. 
Additionally, we avoid close stars being overcorrected by the adopted extinction correction, since most stars with $r>15$ mag are behind the entire dust column~\citep{2008ApJ...673..864J}. On the other hand, with the somewhat conservative faint limit of $r=20.5$ we make sure that only very few stars have a magnitude in the $g$ or $i$ band beyond the 95\% completeness limit of 22.2 mag or 21.3 mag, respectively~\citep[][Table 1]{2007ApJS..172..634A}\footnote{See also \texttt{sdss.org/dr6/}}.

The average seeing of the SDSS imaging data (median PSF width) is $1.4\arcsec$ in the $r$-band~\citep[][]{2008ApJS..175..297A}\footnote{See also \texttt{http://www.sdss.org/dr6/} and the DR5 paper~\citep{2007ApJS..172..634A}}. 
To be on the safe side, we took the minimum angular separation to be $\theta_{\min}=2\arcsec$.

\subsection{Contaminations}
\begin{figure}
   \resizebox{\hsize}{!}{\includegraphics{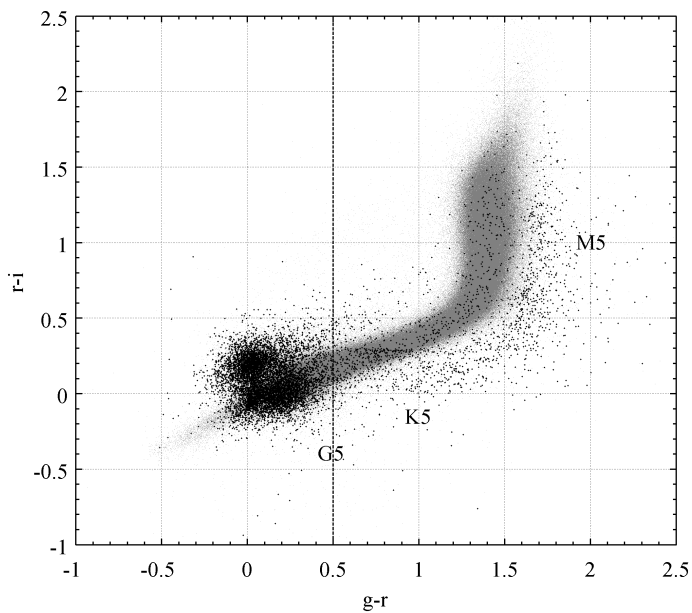}}
      \caption{Colour-colour diagram of all point-like objects within our coordinate's and apparent magnitude's limits (grey points). Black points are SDSS quasar candidates. The vertical dashed line indicates the cut at $g-r=0.5$ mag (see text). Approximate spectral classes are indicated.}
         \label{colcol}
   \end{figure}

Matching our sample with the \emph{QsoBest} table\footnote{See \texttt{cas.sdss.org/dr6/en/help/docs/algorithm.asp?key=\\qsocat}} resulted in the exclusion of 10\,041 quasar candidates. 
Most of them (8\,157) have $g-r\lesssim 0.5$ mag and are scattered in a colour-colour diagram around the otherwise narrow stellar locus as shown in Fig.~\ref{colcol}. 
Even after removing the quasar candidates, the remaining objects in the blue part of our sample show a suspicious scatter, which is probably caused by further quasars and misidentified galaxies. 
We therefore decided to exclude all objects with $g-r<0.5$ mag, removing a further 286\,227 objects from our sample. 
Furthermore, we reject all moving objects (asteroids) by cutting on the {\tt DEBLENDED\_AS\_MOVING} flag. This leaves us with 670\,388 objects classified as ``stars'' in the sample.
 
The large majority of the stars observed by the SDSS are main sequence (MS) stars. 
\citet{2000AJ....120.2615F} estimate the fraction of stars that are not on the MS to be $\sim1.0\%$, most of them giants and subgiants ($\sim90\%$), but also horizontal 
branch stars ($\sim10\%$). 
The number of white dwarfs observed by the SDSS is negligible compared to the number of MS stars~\citep[e.g.][]{2006AJ....131..571H}.
Thus, it seems well-justified to assume that \emph{all} the stars in our sample are on the MS. 

The cut discussed above at $g-r=0.5$ mag implies that our sample contains only stars from spectral type later than about G5 \citep{2000AJ....120.2615F}.
In addition, this cut is appropriate for our purposes for the following three reasons: 

\emph{i)} Because they are very young, the bluest MS stars are mostly members of loose associations, so their clustering properties still represent the peculiarities of their birth places. Being an interesting subject to study \citep[e.g.][]{2007ApJ...670..747K}, excluding them assures that our clustering signal is predominantly due to wide binaries in the field that have lost their memory of their birth places. 

\emph{ii)} As the MS becomes more sparsely populated towards the blue end, and a significant fraction could be made up by metal-poor halo giants, the assumption that all stars in our sample are on the MS might not be valid for the bluest stars. 

\emph{iii)} For magnitudes $M_V\gtrsim 4.5$ mag, the shape of the halo luminosity function agrees well with that of the disc luminosity function  \citep[e.g.][their Fig.~2]{1985ApJ...299..616B}. Thus, the cut on $g-r=0.5$ mag, which corresponds to a cut at $M_V\simeq 5.6$ mag, allows us to use the disc luminosity function for the halo component. 

\subsection{Survey holes and bright stars}\label{hole}
The sample chosen contains some regions where, for various reasons, no object could be observed. These regions are masked\footnote{See \texttt{sdss.org/dr6/algorithms/masks.html}} as \emph{hole} in the SDSS database and can therefore be easily identified. Holes in the sample affect our analysis in two ways: first, they diminish the total area of the sample. Second, they introduce an edge effect, as stars near a hole show a lack of neighbouring stars. The former effect can be easily corrected for in an approximate way as the SDSS provides the radius $\theta_{\rm M}^{(i)}$ of its bounding circle for every mask $i$. 
The residual area of the sample is then
\begin{equation}
\Omega\simeq\Omega_{\rm tot}-\sum_{i=1}^{N_{\rm M}}\pi\left(\theta_{\rm M}^{(i)}\right)^2
\end{equation}
where $N_{\rm M}$ denotes the number of hole masks.
We discuss the correction for edge effects in Sect.~\ref{boundEff} in the context of the correlation function.

Restricting ourselves to the masks defined in the $r$-band, we find $N_{\rm SH}=9$ regions masked as ``hole'' in our sample (survey holes: SH) with an average radius of $\bar{\theta}_{\rm SH}\simeq7.76$ arcmin. These survey holes diminish the total area of the sample by approximately 0.46 square degrees, constituting a marginal correction that could be safely ignored. 

Very bright (saturated) stars cause similar problems: 
a very bright object may appear like a hole in our sample, when the underlying fainter stars blended with this object cannot be revealed. Within our sample there are $N_{\rm BS}=1\,790$ \emph{bright star} (BS) masks in the $r$-band. We exclude all objects inside such a mask, removing 545 stars from our sample. The bounding circles of bright star masks have an average radius of $\bar{\theta}_{\rm BS}\simeq1.95$ arcmin, resulting in a further diminishing of the total area of the sample of circa 5.96 square degrees. 
By correcting the sample's total area for both hole and bright star masks, we get a residual area $\Omega\simeq675-0.46-5.96$ deg$^2=668.58$ deg$^2$. 

% Fig. 2, colcol

\subsection{Final sample}
Our final sample contains $N_{\rm obs}=669\,843$ MS stars with \emph{clean photometry}, a spectral type later than about G5 ($g-r\geq0.5$ mag), and an apparent magnitude in the range $15\leq r\leq 20.5$ mag. The stars are distributed over a solid angle $\Omega\simeq668.58$ square degrees (after correcting for bright star masks and minor survey holes), corresponding to a mean surface density of $n=N_{\rm obs}/\Omega\simeq1\,000$ stars per square degree. 
Throughout this study, we use  the terms ``total sample''  and ``final sample'' synonymously.

%__________________________________________________________________

\section{Stellar correlation function}\label{scf}

\subsection{Estimation of the correlation function}\label{estCF}
The angular two-point \emph{auto}-correlation function (2PCF) $w(\theta)$ is defined via the joint probability $\mbox{d}{\mathcal P}$ of finding objects in both the solid angles $\mbox{d}\Omega_1$ and $\mbox{d}\Omega_2$
\begin{equation}\label{dP}
\mbox{d}{\mathcal P}=n^2\left(1+w(\theta)\right)\mbox{d}\Omega_1\mbox{d}\Omega_2\,,
\end{equation}
where $n$ is the mean density of objects in the sky and $\theta$ is the separation between the two areas~\citep[e.g.][\S45]{1980lssu.book.....P}. 
The number of distinct stellar pairs, $F(\theta)\mbox{d}\theta$, with an angular separation between $\theta$ and $\theta+\mbox{d}\theta$, \emph{observed} in a region of angular size $\Omega$, can then be written as
\begin{equation}\label{F}
F(\theta)\mbox{d}\theta=\frac{N_{\rm obs}(N_{\rm obs}-1)}{2}\left(1+w(\theta)\right)\frac{\mbox{d}\Omega(\theta)}{\Omega}
\end{equation}
where $N_{\rm obs}$ is the number of stars observed in $\Omega$ and $\mbox{d}\Omega(\theta)$ is approximately equal to the solid angle of a ring with (middle) radius $\theta$ and width $\mbox{d}\theta$
\begin{equation}\label{dOm}
\mbox{d}\Omega(\theta)\simeq2\pi\theta\mbox{d}\theta\,.
\end{equation}

Solving for $w$ in Eq.~\ref{F} yields a simple estimator for the 2PCF:
\begin{equation}\label{w}
\hat{w}(\theta)=\frac{F(\theta)}{P(\theta)}-1\,,
\end{equation}
where $P(\theta)$ is the number of pairs expected from a random sample with $N_{\rm obs}$ points ($w=0$ in Eq.~\ref{F}):
\begin{equation}\label{P}
P(\theta)\mbox{d}\theta\simeq\frac{\pi N_{\rm obs}(N_{\rm obs}-1)}{\Omega}\theta\mbox{d}\theta\,.
\end{equation}
The 2PCF estimate $\hat{w}(\theta)$ is a measure for the excess of observed pairs separated by an angle $\theta$ with respect to a randomly distributed sample.

Another statistical measure equivalent to the 2PCF, which we use for visualising the data, is the 
\emph{cumulative difference distribution} (CDD) $\hat\gamma(\theta)$, giving the cumulative number of pairs in excess of a random distribution:
\begin{equation}\label{CDD}
\hat\gamma(\theta)\equiv\int_{\theta_{\min}}^\theta\left(F(\theta')-P(\theta')\right)\mbox{d}\theta'=\int_{\theta_{\min}}^\theta\hat{w}(\theta')P(\theta')\mbox{d}\theta' \,,
\end{equation}
with $\theta_{\min}=2\arcsec$ (see above).
The CDD is completely equivalent to the 2PCF, but is -- as for all cumulative distributions -- more sensitive to statistical trends caused over a wider range of angular separations (i.e.~over several bins), at the expense of a strong correlation of its values at different separations.
The use of the simple estimator in Eq.~\ref{w} is appropriate as long as boundary effects due to the finite sample size are negligible \citep{1994ApJ...424..569B}. We discuss boundary effect in the next section.
The number of pairs observed $F(\theta)$ is determined very efficiently \emph{inside} the SDSS database using the precalculated \emph{Neighbors} table which contains all the objects within $\theta_{\max}=30\arcsec$ of any given object.

\subsection{Boundary effects}\label{boundEff}
Stars close to the sample's boundary have a somewhat truncated ring $\mbox{d}\Omega(\theta)$ and consequently show a lack of neighbours. 
However, as in the present study the probed angular scale is small compared to the sample's size, it turns out that these edge effects have a negligible impact on our results: we estimate the relative error introduced in $\mbox{d}\Omega(\theta)$ by omitting the edge correction to be less than $\sim 0.04\%$ for any given $\theta\leq 30\arcsec$. 

The stars near the boundary of a hole or a bright star mask (in the following simply ``hole'') show a lack of neighbours, too. Therefore, only a fraction $F_{\rm H}^{\rm tot}(\theta)$ of all pairs separated by an angular distance $\theta$ has been observed, and we need to correct the observed number of pairs $F(\theta)$. The ``true'' number of pairs, corrected for edge effects, then reads as
\begin{equation}
F_{\rm corr}(\theta)=\frac{F(\theta)}{F_{\rm H}^{\rm tot}(\theta)}\,.
\end{equation}
In calculating the fraction $F_{\rm H}^{\rm tot}$ we proceeded in the same way as~\citet{1998MNRAS.301..289L}. 
The calculation is outlined in Appendix \ref{EC4H}. We find that the relative error in $F$ for $\theta=30\arcsec$ amounts to
\begin{equation}
\Delta_F\equiv\frac{F_{\rm corr}-F}{F_{\rm corr}}=1-F_{\rm H}^{\rm tot}\simeq0.7\%\,.
\end{equation}
For smaller angular separations, the effect is even less.

Taking into consideration these edge effects, we rewrite the 2PCF estimate
\begin{equation}\label{wcorr}
\hat{w}_{\rm corr}(\theta)=\frac{F_{\rm corr}(\theta)}{P(\theta)}-1
\end{equation}
and the CDD
\begin{equation}\label{CDDcorr}
\hat\gamma_{\rm corr}(\theta)=\int_{\theta_{\min}}^\theta\hat{w}_{\rm corr}(\theta')P(\theta')\mbox{d}\theta' \,.
\end{equation}
  
\subsection{Uncertainty of the correlation function estimate}\label{uncert}
As the values of $\hat{w}_{\rm corr}$ at different separations are \emph{not} independent, Poisson errors may underestimate the uncertainty in $\hat{w}_{\rm corr}$, especially when the angular scales under consideration are large~\citep{1993ApJ...417...19H,1994ApJ...424..569B}.
However, as we see in Sect.~\ref{results}, the clustering of wide binary stars occurs on small angular scales ($\theta\leq15\arcsec$) compared to the size of our sample (20 deg $\times$~40 deg). We therefore expect that using Poisson errors only underestimates the true errors by a small amount in our case. 
Thus, we adopt Poissonian errors on $F(\theta)$:
\begin{equation}\label{dF}
\delta F(\theta)=\sqrt{F(\theta)}\quad\mbox{and}\quad\delta F_{\rm corr}(\theta)=\sqrt{\frac{F_{\rm corr}(\theta)}{F_{\rm H}^{\rm tot}(\theta)}}\,. 
\end{equation}
Using Eq.~\ref{wcorr} and Gauss' error propagation formula, we may write the uncertainty of the CF estimate $\hat{w}_{\rm corr}$ as
\begin{equation}\label{dw}
\delta\hat{w}_{\rm corr}=\frac{\delta F_{\rm corr}}{P}=\sqrt{\frac{\hat{w}_{\rm corr}+1}{F_{\rm H}^{\rm tot}P}}\,,
\end{equation}
where we omitted the dependencies on $\theta$ for the sake of brevity.

The uncertainty in the CDD $\hat\gamma$ is easily obtained in the same way:
\begin{equation}\label{dCDD}
\delta\hat\gamma_{\rm corr}(\theta)=\sqrt{\int_{\theta_{\min}}^\theta\delta F_{\rm corr}^2(\theta')\mbox{d}\theta'}\,.
\end{equation}

\subsection{Testing the procedure for a random sample}
% Fig. 3, ranSam (new)
   \begin{figure}
   \resizebox{\hsize}{!}{\includegraphics{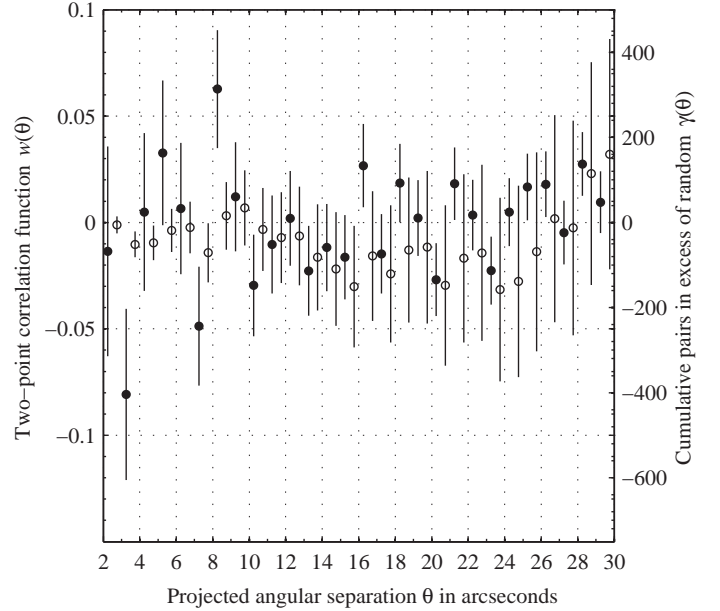}}
      \caption{2PCF as inferred from a random sample (solid circles, left ordinate) and the corresponding CDD (open circles, right ordinate). Poisson errors are indicated as vertical lines.}
         \label{ranSam}
   \end{figure}

To test the validity of the procedure to estimate the 2PCF described in the previous sections, we generated a random sample having the same number of ``stars'' and distributed over the same area as the stars in our final sample. In addition, we made certain that the random sample contains the same number of holes with appropriate radii.

For the analysis of the random sample we wrote a dedicated computer program that lays a grid with a mesh size of 1 arcmin over the sample before counting the pairs. This grid acts as a distance filter and avoids calculating the distances between all possible pairs: only pairs with both their components within a cell or with them in two adjacent cells, respectively, were taken into account.

If our procedure is correct we would expect a zero signal in both the 2PCF and the CDD when analysing a sample of randomly distributed stars. Figure \ref{ranSam} shows the result of the analysis of our random sample. There is indeed no clustering signal evident, neither in the 2PCF nor in the CDD. The results are consistent throughout with a zero signal out to the separation limit of 30\arcsec  (compare also with Fig.~\ref{allModsTot_b}). This shows that our procedure for estimating the 2PCF is reliable.

%__________________________________________________________________

\section{The model}\label{model}
Our approach to modelling the angular 2PCF is based on a technique developed by \citet*{1987ApJ...312..390W},\defcitealias{1987ApJ...312..390W}{WW87} hereafter \citetalias{1987ApJ...312..390W}. As we describe in more detail in the following sections, this technique
makes some simple assumptions on the basic statistical properties of wide binaries,
and projects these theo\-retical distributions on the observational plane using the selection criteria of a given (binary) star catalogue and the geometry of the Milky Way galaxy \citep*[see also][]{1988Ap&SS.142..277W}. 

Their long periods make it virtually impossible to distinguish wide binary stars from mere
chance projections (optical pairs) by their orbital motion. Therefore, we do not attempt to identify individual wide 
binaries, but we look for a \emph{statistical} signal stemming from physical wide pairs in the sample, solely exploiting precise stellar position measurements provided by the SDSS. 

The original \citeauthor{1987ApJ...312..390W} (WW) technique calculates the projected separation distribution to compare it with a sample of wide binaries of known distance and angular separation. 
In the present study we are only dealing nwith angular separations. The calculation of wide binary counts as a function of angular separation
alone requires an appropiate modification of the WW-technique \citep[see also][]{1991PhDT.........8G}, which we discuss in Sect.~\ref{WWmod}.

\subsection{Wasserman-Weinberg technique}
\citetalias{1987ApJ...312..390W} developed ``a versatile technique for comparing wide binary observations with theoretical semimajor axis distributions''. 
According to \citetalias{1987ApJ...312..390W} we may write the number of observed binaries $\psi(s)\mbox{d}s$ with projected physical
separations between $s$ and $s+\mbox{d}s$ in a given catalogue of stars as
\begin{equation}\label{psis}
\psi(s)\,\mbox{d}s=n_{\rm WB}Q(s)V(s)\,\mbox{d}s\,,
\end{equation}
where $n_{\rm WB}$ is the total number density of wide binaries in the solar neighbourhood\footnote{We use the term ``solar neighbourhood'' for local quantities.}, $Q(s)$ is the reduced distribution of projected separations, and $V(s)$ the ``effective volume''. 

The total number density $n_{\rm WB}$ is one of the two free para\-meters in the model that will be inferred by fitting the model to the observational data. 
The reduced separation distribution $Q(s)$ contains the physical properties from the wide binaries (semi-major axis distribution, distribution of eccentricities, orientation of the orbital planes) projected onto the observational plane, 
whereas the effective volume $V(s)$ takes into account the characteristics of the sample under consideration (covered area, range of angular separation examined, magnitude limits), as well as the stellar density distribution in the Galaxy and the lumino\-sity function. 

As we show in Sect.~\ref{WWmod}, this neat formal splitting in physical properties and selection effects will not be possible anymore after the required modification on the WW-technique mentioned above. 
At this point, we discuss the two parts -- $Q(s)$ and $V(s)$ -- more in detail.

\subsubsection{Reduced separation distribution}\label{redSepDis}
The reduced separation distribution $Q(s)$ is given by the projection of the reduced (present-day) semi-major axis distribution $q(a)$ against the sky.
Gravitational perturbations due to stars, giant molecular clouds, and (hypothetical) DM particles cause the semi-major axes of (disc) wide binaries to evolve. Little is known about the initial semi-major axis distribution, and usually a single power-law is assumed, because of its simplicity \citep{1988Ap&SS.142..277W} but also because of theoretical considerations \citep[][and references therein]{1997CeMDA..68...27V,2007IAUS..240..417P}. The evolution of the semi-major axis distribution of disc wide binaries has been modelled by \citet*{1987ApJ...312..367W} in terms of the Fokker-Plack equation. Their numerical simulations suggest that the semi-major axis distribution evolves in a self-similar way for reasonable choices of the initial power-law index and the wide binary birth rate function. 
We decided, therefore, to model the present-epoch semi-major axis distribution of wide binary stars $q(a)$ by a single powerlaw 
\begin{equation}\label{qa}
q(a)\propto\left(\frac{a}{\mbox{pc}}\right)^{-\lambda}~\mbox{pc}^{-1}
\end{equation}
with the power-law index $\lambda$, the second free parameter of the model.
The range in $a$ where Eq.~\ref{qa} is a valid description of the observed semi-major axis distribution at the same time provides the \emph{definition} of what we consider as a ``wide binary star''. We specify this range by an upper and a lower limit, $a_{\min}$ and $a_{\max}$, respectively.
There is observational evidence that the distribution of the semi-major axis of binary stars has a break at $0.001$ pc~\citep{1983ARA&A..21..343A}. 
Following \citetalias{1987ApJ...312..390W} and~\citet{1988Ap&SS.142..277W}, we thus take the lower limit, providing the division between the close and wide binary populations, to be $a_{\min}=0.001$ pc ($\approx200$ AU). 
On the other hand, the Galactic tides provide a natural maximum semi-major axis $a_{\rm T}$ beyond which no bound orbits can exist. Details on the calculation of $a_{\rm T}$ are given in Appendix \ref{aT}. We find it to be of the order of 1 pc.

We normalise $q(a)$ -- hence the name ``\emph{reduced} semi-major axis distribution'' -- such that
\begin{equation}
\int_{a_{\min}}^{a_{\rm T}}q(a)\,\mbox{d}a=1
\end{equation}
giving
\begin{equation}
q(a)=c_\lambda\left(\frac{a}{\mbox{pc}}\right)^{-\lambda}~\mbox{pc}^{-1}
\end{equation}
where
\begin{equation}
c_\lambda=\left\{\begin{array}{ll} (1-\lambda)\left[a_{\rm T}^{1-\lambda}-a_{\min}^{1-\lambda}\right]^{-1}, & \mbox{for }\lambda\neq 1\\\\ 
\left[\ln a_{\rm T}-\ln a_{\min}\right]^{-1}, & \mbox{for }  \lambda=1\end{array}\right. 
\end{equation}
with $a_{\min}$ and $a_{\rm T}$ in pc.

In a more general treatment, $q(a)$ might also depend on the luminosity classes of the binary star members, as well as on their magnitudes. Like \citetalias{1987ApJ...312..390W}, we restrict our analysis to models where $q(a)$ only depends on the 
semi-major axis $a$. 

It is somewhat disputed whether the semi-major axis distribution has a break at larger $a$ attributed to the disruptive effects of the environment on the widest binary stars. In their extensive work \citet{1991ApJ...382..149W}~\citep[see also][]{1990bdm..proc..117W} conclud that ``although the data \emph{suggest} a break in the physical distribution of wide binary separations, they do not \emph{require} a break with overwhelming statistical significance'', whereas the more recent study by \citet{2007AJ....133..889L} shows evidence for a break at $s\sim3\,500$ AU (statistically, we have $\langle s\rangle\simeq a$; see Eq.~\ref{sEQa}). 
In this context, the question arises whether the assumption that the data is described by a single powerlaw (Eq.~\ref{qa}) up to the tidal limit $a_{\rm T}$ can be rejected with confidence. We address this question in Sect.~\ref{results}.

Assuming that ``the binaries' orbital planes are randomly oriented and that their eccentricities $e$ are 
distributed uniformly in $e^2$'' (\citetalias{1987ApJ...312..390W})
we can project $q(a)$ against the sky by \citep{1988Ap&SS.142..277W}
\begin{equation}\label{Qs}
Q(s)=\int_{s/2}^{\infty}\frac{\mbox{d}a}{a}q(a){\mathcal F}(s/a)\,,
\end{equation}
where ${\mathcal F}(x)$ takes the eccentricity and angular averaging into account for the pairs, which may be written as
\citep[\citetalias{1987ApJ...312..390W},][]{1988ApJ...329..253W}
\begin{equation}\label{fx}
{\mathcal F}(x)=\frac{4x}{\pi}\int_0^{\sqrt{2-x}}\mbox{d}u\sqrt{\frac{(u^2+x)(2-x-u^2)}{u^2+2x}}\,.
\end{equation}

With the above assumptions on the orientations of the orbital planes and the distribution of eccentricities it can be shown that the \emph{average} observed projected separation $\langle s\rangle$ for a given semi-major axis $a$ is basically equal to $a$ \citep[e.g.][]{2004ApJ...601..311Y}:
\begin{equation}\label{sEQa}
\langle s\rangle=\frac{5\pi}{16}a\simeq0.98a\,.
\end{equation}
From the normalisations of $q(a)$ it also follows that $Q(s)$ is normalised to unity in the range $[\langle s\rangle_{\min},\langle s\rangle_{\max}]$.

Substituting $\eta=\sqrt{s/2a}$ in Eq.~\ref{Qs}, as well as $\xi=u/\sqrt{2}$ in Eq.~\ref{fx}
leads to a convenient form for numerical integration
\begin{equation}
Q(s)=\left(\frac{s}{\mbox{pc}}\right)^{-\lambda}C_{\lambda}(s)\,~\mbox{pc}^{-1},
\end{equation}
where
\begin{equation}\label{Cs}
C_{\lambda}(s)=2^{\lambda+5}\,\frac{c_{\lambda}}{\pi}\hspace{-0.8cm}\int\limits_{\min\left(1,\,\sqrt{s/2a_{\rm T}}\right)}^
{\min\left(1,\,\sqrt{s/2a_{\min}}\right)}\hspace{-0.8cm}\mbox{d}\eta\,\eta^{2\lambda+1}\!
\hspace{-0.2cm}\int\limits_0^{\sqrt{1-\eta^2}}\hspace{-0.2cm}\mbox{d}\xi\sqrt{\frac{(\xi^2+\eta^2)(1-\eta^2-\xi^2)}
{\xi^2+2\eta^2}}\,.
\end{equation}
The integrals in Eq.~\ref{Cs} are evaluated using Gauss quadrature
\citep{1992nrca.book.....P}.
For $s$ far from $a_{\min}$ or $a_{\rm T}$ the distribution of the projected separations $Q(s)$ is approximately powerlaw
\begin{equation}
Q(s)\propto\left(\frac{s}{\mbox{pc}}\right)^{-\lambda}\,~\mbox{pc}^{-1}.
\end{equation}
 
\subsubsection{Effective volume}\label{Veff}
% Fig. 4, effVol_galMod
   \begin{figure}
   \resizebox{\hsize}{!}{\includegraphics{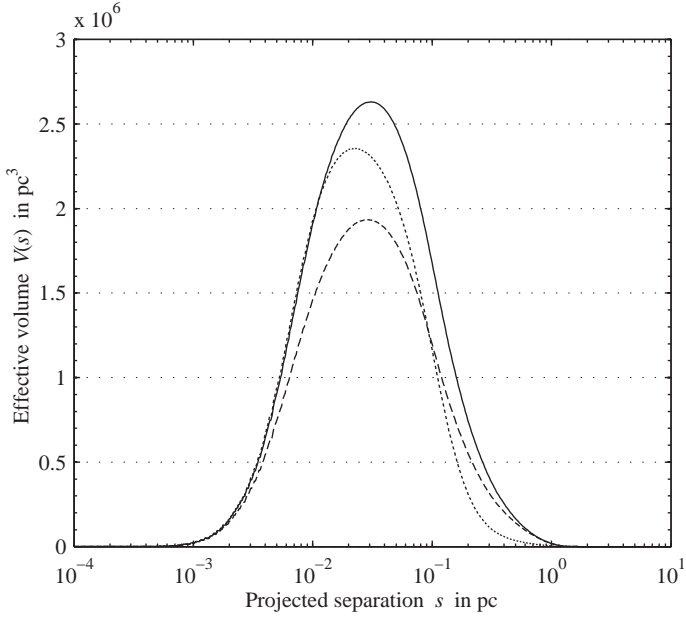}}
      \caption{Effective volumes calculated using the parameters that correspond to our final sample (Table \ref{subsamParam}) for the three Galactic structure parameter sets as discussed in \S\ref{galmod}. Long dashed line: set 1; solid line: set 2; short dashed line: set 3.}
         \label{effVoll_galMod}
   \end{figure}
 
As we deal with a magnitude-limited sample, we are plagued by selection effects that must be properly taken into account. Following \citetalias{1987ApJ...312..390W}, we do this by means of the \emph{effective volume}. It allows for the solid angle $\Omega$ covered by the sample, the angular separation range $[\theta_{\min},\theta_{\max}]$ we are examining, and the apparent magnitude limits $m_{\min}$ and $m_{\max}$. Furthermore, the effective volume takes the stellar density distribution $\rho$ into account, as well as the normalised (single star) luminosity function $\widetilde\Phi$ (both described more in detail in Sect.~\ref{galmod}).

``Making the plausible assumptions that the intrinsic luminosities of the two stars in a wide binary are independent and distributed in the same way as the luminosities of field stars'' (\citetalias{1987ApJ...312..390W}), and assuming further that the stellar density distribution $\rho$ is independent of the stars' absolute luminosities, we may write the effective volume as
\begin{equation}\label{Vs}
V(s)=\Omega\hspace{-0.2cm}\int\limits_{s/\theta_{\max}}^{s/\theta_{\min}}\hspace{-0.2cm}\mbox{d}DD^2
\,\tilde\rho(D)\hspace{-0.2cm}\iint\limits_{M_{\min}(D)}^{M_{\max}(D)}\hspace{-0.2cm}\mbox{d}M_1\mbox{d}M_2
\,\widetilde\Phi(M_1)\widetilde\Phi(M_2)\,,
\end{equation}
where we have the standard relations  
\begin{eqnarray}
M_{\max}(D)&=&m_{\max}-5\log_{10}(D/10~\mbox{pc})\label{Mmin}\\
M_{\min}(D)&=&m_{\min}-5\log_{10}(D/10~\mbox{pc})\label{Mmax}\,,
\end{eqnarray}
and $\tilde\rho(D)$ is normalised such that $\tilde\rho(0)=1$.

In Fig.~\ref{effVoll_galMod} the effective volumes for the three Galactic structure parameter sets (described in the next section) are plotted using the parameters that correspond to our final sample (see Table \ref{subsamParam}). 
From the effective volume $V(s)$, it is evident that our study is most sensitive to projected separations in $0.01\mbox{ pc}\la s\la 0.1$ pc, whereas our study is absolutely insensitive when $s\la 0.001$ pc or $s\ga 1$ pc. 
Referring to Eq.~\ref{sEQa}, we see that the shape of the effective volume also indicates that we are insensitive to semi-major axis $a\la 0.001$ pc and for $a\ga 1$ pc, which nicely fits the range in $a$ where we assume that the single power-law model holds. But it also shows that our analysis is not too sensitive to very wide binary stars with semi-major axes larger than 0.1 pc.

\subsection{Galactic model}\label{galmod}
\subsubsection{Stellar density distribution}\label{rhoD}

Following earlier work \citep[especially][and references therein]{2008ApJ...673..864J,2001ApJ...553..184C}, we modelled the stellar density distribution of the Milky Way Galaxy by including two exponential disc components -- a thin and a thick disc -- and an elliptical halo component whose density profile obeys a powerlaw. 
The contribution of the Galactic bulge is negligible in the direction of the NGP and it is therefore ignored in the following. 

The three components are added to yield the total density distribution 
\begin{equation}
\rho(D)=\rho(0)\tilde\rho(D)
\end{equation}
with
\begin{equation}
\tilde\rho(D)=\frac{\tilde\rho^{\rm thin}(D)+n^{\rm thick}\tilde\rho^{\rm thick}(D)+n^{\rm halo}\tilde\rho^{\rm halo}(D)}{1+n^{\rm thick}+n^{\rm halo}}\,,
\end{equation}
which is normalised such that $\tilde\rho(0)=1$.
The thick disc and the halo component are normalised with respect to the thin disc via the normalisation constants $n^{\mbox{\tiny thick}}$ and $n^{\mbox{\tiny halo}}$, respectively. The overall normalisation $\rho(0)$ is calibrated to produce the observed star counts.

Both the thin and the thick disc populations obey a double exponential density law of the form~\citep*[e.g.][]{1980ApJS...44...73B}
\begin{equation}\label{rhodisk}
\exp\left(-\frac{r-r_0}{h_r}\right)\exp\left(-\frac{|z|}{h_z}\right)
\end{equation}
where $h_r$ and $h_z$ are the scale length and the scale height of the disc, $z$ is the object's height above the Galactic midplane, and $r$ is its Galactocentric distance in the Galactic plane. 
We take the Sun's distance from the centre of the Galaxy in the Galactic plane to be $r_0=8$ kpc. 
\citet{2008ApJ...673..864J} find that the Sun is located $z_0\simeq 25$ pc above the Galactic midplane (\citet{2001ApJ...553..184C} give $z_0\simeq 27$ pc). 
It is straightforward to derive the following useful relations
\begin{eqnarray}
z&=&z_0+D\sin b\\
r^2&=&r_0^2+d^2-2r_0d\cos\ell\\
d&=&D\cos b\,,
\end{eqnarray}
where $d$ is its distance from the Sun in the Galactic plane and $\ell$ and $b$ are its Galactic longitude and latitude, respectively.
The stellar density is fairly constant within the sample, so we neglect the direction-dependent density variations. For the sake of simplicity, we adopt the coordinates of our sample's centre: $\ell=175.6^{\circ}$ and $b=81.6^{\circ}$. (Dividing our sample into subsamples and summing over them taking their centres hardly influences our results.)

In line with \citet{2008ApJ...673..864J} and~\citet{2001ApJ...553..184C} we assume the scale heights to be independent of absolute magnitude $M$.
Using capitals for the thick disc's and lower case letters for the thin disc's scale height and length, we may write the normalised number density distribution of the stars in the Galactic thin disc as
\begin{equation}
\tilde\rho^{\rm thin}(D)=\exp\left(-\frac{r-r_0}{h_r}-\frac{|z|}{h_z}\right)\cdot\exp\left(\frac{z_0}{h_z}\right)
\end{equation}
and that of the thick disc as
\begin{equation}
\tilde\rho^{\rm thick}(D)=\exp\left(-\frac{r-r_0}{H_r}-\frac{|z|}{H_z}\right)\cdot\exp\left(\frac{z_0}{H_z}\right)\,,
\end{equation}
where the rightmost factors are for normalisation purposes and ensure that $\tilde\rho^{\rm thin}(0)=\tilde\rho^{\rm thick}(0)=1$.

The density distribution of the stellar halo population is modelled by a powerlaw with index $k$. The observational data prefer a somewhat oblate halo giving an ellipsoid, flattened in the same sense as the Galactic disc, with axes $a=b$ and $c=\kappa a$, where $\kappa$ controls the ellipticity of the halo \citep[cf.][]{2008ApJ...673..864J}
\begin{equation}\label{rhohalo}
\tilde\rho^{\rm halo}(D)=\left[\frac{r^2+\left(\frac{z}{\kappa}\right)^2}{r_0^2+\left(\frac{z_0}{\kappa}\right)^2}\right]^{-\frac{k}{2}}\,,
\end{equation}
which, of course, also satisfies $\tilde\rho^{\rm halo}(0)=1$.

We compare three different sets of structure parameters:

{\it Set 1:} The \emph{measured} values from \citet{2008ApJ...673..864J} (see their Table 10). These values are best-fit parameters as \emph{directly} measured from the apparent number density distribution maps using their ``bright'' photometric parallax relation (their Eq.~2). They are not corrected for bias caused by, e.g., unresolved stellar multiplicity, hence the term ``apparent''.

{\it Set 2:} The \emph{bias-corrected} values from \citet{2008ApJ...673..864J} (see again their Table 10). 
These values were corrected for unrecognised stellar multiplicity, Malmquist bias, and systematic distance determination errors by means of Monte Carlo-generated mock catalogues \citep[][\S4.3.]{2008ApJ...673..864J}. Following \citet{1997AJ....113.2246R}, the fraction of ``stars'' in the local stellar population that in fact are unresolved binaries is taken to be 35\%. 
However, the halo component was not included in the mock catalogues, and its structure parameters were therefore not corrected, but the measured values are used instead.
This set of parameters will be referred to as our standard set.

{\it Set 3:} As a third independent set of Galactic structure parameters, we refer to the somewhat earlier work of \citet{2001ApJ...553..184C}. It is also based on observations obtained with the SDSS, but \citet{2001ApJ...553..184C} infer the density distribution of the stars by inverting the fundamental equation of stellar statistics \citep[e.g.][]{1996fuas.book.....K}.

% table 1, strucParam 
\begin{table}
\caption{Galactic structure parameters}            
\label{strucParam}      
\centering
\renewcommand{\footnoterule}{}                   
\begin{tabular}{llll}       
\hline\hline                 
structure & set 1 & set 2 & set 3 \\
parameter & ~ & ~ & ~ \\ 
\hline                       
   $z_0$ \, \, [pc] & 25 & 25 & 27 \\      
   $h_z$ \, \, [pc] & 245 & 300 & 330 \\
   $h_r$ \, \, [pc] & 2\,159 & 2\,600 & 2\,250 \\
   $n^{\rm thick}$ &  0.13 & 0.12 & 0.0975 \\
   $H_z$ \,\, [pc] &  743 & 900 & 665 \\
   $H_r$ \,\, [pc] &  3\,261 & 3\,600 & 3\,500 \\
   $n^{\rm halo}$ &  0.0051 & 0.0051 & 0.00125 \\
   $\kappa$ & 0.64 & 0.64 & 0.55 \\
   $k$ & 2.77 & 2.77 & 2.5 \\
\hline                                   
\end{tabular}
\end{table}

These three sets of Galactic structure parameters are summarised in Table \ref{strucParam}. \citet{2008ApJ...673..864J} quote  unrecognised multiplicity as one of the dominant sources of error in the distance determination of the stars. (Only the uncertainties in absolute calibration of the photometric parallax relation might be even more important, but little can be done to increase its accuracy at the moment; see, however, \citet{2008ApJ...689.1244S}.) It is therefore not surprising that the scale heights and lengths of the bias-corrected parameter set are larger than those of the measured parameters as the misidentification of a binary star as a single star results in an underestimation of its distance. Regarding the values derived by \citet{2001ApJ...553..184C}, we find the largest difference in the normalisation of the halo component that is about four times smaller than that given by \citet{2008ApJ...673..864J}, whereas the other parameters are not too dissimilar.

\subsubsection{Stellar luminosity function}\label{sec:LF}
For our study we refer to the luminosity function (LF) inferred by \citet{1997ESASP.402..675J} using Hipparcos parallaxes, which gives reliable values for a wide magnitude range ($-1\leq M_V\leq19$). (The faint end of the LF is somewhat uncertain and \citet{1997ESASP.402..675J} give only a lower limit in the range $20\leq M_V\leq23$. We take that lower limit at the faint end to be the true value of the LF in that magnitude range.)

We need to transform the  \citet{1997ESASP.402..675J} LF from visual ($V$-band) into $r$-band magnitudes, i.e.~from $\Phi(M_V)$ to $\Phi(M_r)$. We perform this transformation in an unsophisticated way by combining the photometric parallax relation from \citet{1988AJ.....95.1843L}, which was calibrated using the Hyades and is linear in $(B-V)$ (their Eq.~1a)
\begin{equation}
M_V=5.64(B-V)+1.11~\mbox{mag}
\end{equation}
with a transformation equation that is obtained by subtracting Eq.~1 from 3 in \citet{2005AN....326..321B}
\begin{equation}
V-r=0.491(B-V)-0.144~\mbox{mag}\,.
\end{equation}
The linearity of the photometric parallax relation removes any inversion problem, since it assures -- in a mathematical sense -- that a colour index $(B-V)$ exists for every $M_V$.
For a given distance we then have
\begin{equation}
M_r=0.913M_V+0.0476~\mbox{mag}
\end{equation}
and the transformed LF is simply
\begin{equation}
\Phi(M_r)=\Phi(M_V)\frac{\mbox{d}M_V}{\mbox{d}M_r}\,.
\end{equation}
Using linear interpolation (the brightest bin at $M_r=-1$ mag is an \emph{extra}polation), we finally have rebinned $\Phi(M_r)$ such that the bins are centred on integer values of absolute magnitude $M_r$ and have a width of $\Delta M_r=1$ mag.

We normalise the LF by integrating over the absolute magnitude range corresponding to our total sample
\begin{equation}
\widetilde\Phi(M_r)=\frac{\Phi(M_r)}{\int_{5.2}^{\infty}\Phi(\mu)\mbox{d}\mu}=\frac{\Phi(M_r)}{n'_*}
\end{equation}
with $n'_*\simeq0.094~\mbox{pc}^{-3}$ being the total number density of stars in the solar neighbourhood having $g-r > 0.5$ mag. Assuming that all stars are on the MS, we find, using the (``bright'') photometric parallax relation from \citet{2008ApJ...673..864J}, that the cut at $g-r=0.5$ mag corresponds to $M_r\simeq5.2$ mag. (We first transformed it from $g-r$ into $r-i$ using Eq.~4 from \citeauthor{2008ApJ...673..864J}) 

\subsection{Modification of the Wasserman-Weinberg technique}\label{WWmod}
To derive a theoretical angular 2PCF we need to calculate the expected number of wide 
binaries as a function of \emph{angular} separation. To this end, the above-described WW-technique requires the modification that we address now.

Let $\varphi(\theta)\mbox{d}\theta$ be the number of wide binaries 
observed with an angular separation between $\theta$ and $\theta+\mbox{d}\theta$. For a given distance $D$, we have 
$\varphi(\theta)\mbox{d}\theta=\psi(s)\mbox{d}s$ and $s=D\theta$. (Note that our unit for $s$ and $D$ is pc, 
whereas $\theta$ is in rad.)
The latter expression can be used to write the reduced separation distribution $Q(s)$ as $Q(D\theta)$. This introduces an explicit dependency on $D$, so we need to incorporate $Q(D\theta)$ into the integration over $D$ in the formula for the effective volume (Eq.~\ref{Vs}).
This incorporation is the reason it is no longer possible to separate the reduced separation distribution 
from the effective volume in a formal way like in Eq.~\ref{psis}. 

Furthermore, we need to modify the limits in the integration over $D$ in Eq.~\ref{psis}. Recalling that for a given semi-major axis $a$, the average observed projected separation is $\langle s\rangle=0.98a$ (Eq.~\ref{sEQa}), it appears to be appropriate to let the integration limits run from $\langle s\rangle_{\min}/\theta$ to $\langle s\rangle_{\max}/\theta$ \citep{1991PhDT.........8G}.
We take $\langle s\rangle_{\min}=0.98a_{\min}=9.8\cdot10^{-4}$ pc and $\langle s\rangle_{\max}=0.98a_{\rm T}$ (for the calculation of the tidal limit $a_{\rm T}$ see Appendix \ref{aT}).

Putting it all together, we may write the number of observed wide binaries as a function of
angular separation as
\begin{equation}\label{phiTheta}
\varphi(\theta)= n_{\rm WB}\Omega\hspace{-0.3cm}
\int\limits_{\langle s\rangle_{\min}/\theta}^{\langle s\rangle_{\max}/\theta}\hspace{-0.3cm}\mbox{d}DD^3\tilde\rho(D)Q(D\theta)\hspace{-0.3cm}\iint\limits_{M_{\min}(D)}^{M_{\max}(D)}\hspace{-0.3cm}\mbox{d}M_1\mbox{d}M_2
\widetilde\Phi(M_1)\widetilde\Phi(M_2)\,,
\end{equation}
where we have included an additional factor $D$ into the integration over $D$, because $\mbox{d}s=D\mbox{d}\theta$ for a given $D$.

The model-2PCF is now determined by adding the number of physical pairs, calculated by Eq.~\ref{phiTheta}, to the number of pairs expected from a random sample given by Eq.~\ref{P}
\begin{equation}\label{omegaMod}
w_{\mbox{\tiny mod}}(\theta)=\frac{\varphi(\theta)+P(\theta)}{P(\theta)}-1=
\frac{\Omega\,\varphi(\theta)}{\pi N_{\rm obs}(N_{\rm obs}-1)\theta}\,.
\end{equation}
The two free parameters -- $n_{\rm WB}$ and $\lambda$ -- are determined in a least-square sense by fitting the measured 2PCF $\hat{w}_{\rm corr}(\theta)$ to the model-2PCF $w_{\mbox{\tiny mod}}(\theta)$ as described more in detail in the next section.

%__________________________________________________________________

\subsection{Fitting procedure}
We determine the two free parameters of the model, $n_{\rm WB}$ and $\lambda$, by means of a Leven\-berg-Marquardt nonlinear least-square algorithm \citep{lourakis04LM}\footnote{We use \emph{levmar-2.2}.}, which minimises the value of the $\chi^2$ defined as
\begin{equation}\label{chi2}
\chi^2=\sum_{i=1}^N\left(\frac{\hat{w} _{\rm corr}(\theta_i)-w_{\rm mod}(\theta_i)}{\delta\hat{w} _{\rm corr}(\theta_i)}\right)^2\,.
\end{equation}
The data is binned in steps of $\Delta\theta\equiv\theta_{i+1}-\theta_i=1\arcsec$, where $\theta_1=\theta_{\min}=2\arcsec$ and $\theta_N=\theta_{\max}=30\arcsec$ as imposed by the resolution limit of the SDSS and maximal distance in the \emph{Neighbors} table, respectively. As a result, we have $N=28$ here.
Given the approximative character of our study, the use of the this standard definition of the $\chi^2$ is appropriate, even though the values of the 2PCF at different angular separations are not strictly independent of each other.

Following \citet{1992nrca.book.....P}, we use the incomplete gamma function $Q(\chi^2|\nu)$ with $\nu=N-2=26$ degrees of freedom as a quantitative measure of the goodness-of-fit. (N-2 because the model has two free parameters: $n_{\rm WB}$ and $\lambda$.) Values of $Q$ near unity indicate that the model adequately represents the data. 

\subsection{Confidence intervals}
To estimate the uncertainties of our best-fit values, we use Monte Carlo confidence intervals (MCCRs)~\citep[e.g.][\S15.6]{1992nrca.book.....P}. 
Assuming Poissonian errors, we generate $10\,000$ \emph{synthetic} data sets by drawing the number of unique pairs ``observed'' in the $k$-th synthetic sample $F_{\rm syn}^{(k)}(\theta)$ from a Poisson distribution with mean $F(\theta)$, where $F(\theta)$ is the observed number of pairs, \emph{not} corrected for edge effects due to survey holes. The ``true'' number of pairs in a synthetic sample is then given by dividing $F_{\rm syn}^{(k)}$ by $F_{\rm H}^{\rm tot}$ (see \S\ref{boundEff}).

For each synthetic sample, we determine best-fit values for the model's free parameters, $n_{\rm WB}^{(k)}$ and $\lambda^{(k)}$, in a least-square sense as described above. 
A $p\%$-MCCR is defined by the line of constant $\chi^2$ which encloses $p\%$ of the best-fit values in the $n_{\rm WB}$ versus $\lambda$ plane (see Fig.~\ref{MCCR_tot_juric_c}). The confidence intervals of $n_{\rm WB}$ and $\lambda$ are then given by the orthogonal projection of the MCCR onto the corresponding axis.

We include in our error estimate only the uncertainties stemming from the pair counts in $F(\theta)$. Neither the uncertainties in the Galactic structure parameters nor those in the LF are taken into account.

\section{Results}\label{results}
\subsection{Analysis of the total sample}
% Fig. 5, allModsTot_b
   \begin{figure}
   \resizebox{\hsize}{!}{\includegraphics{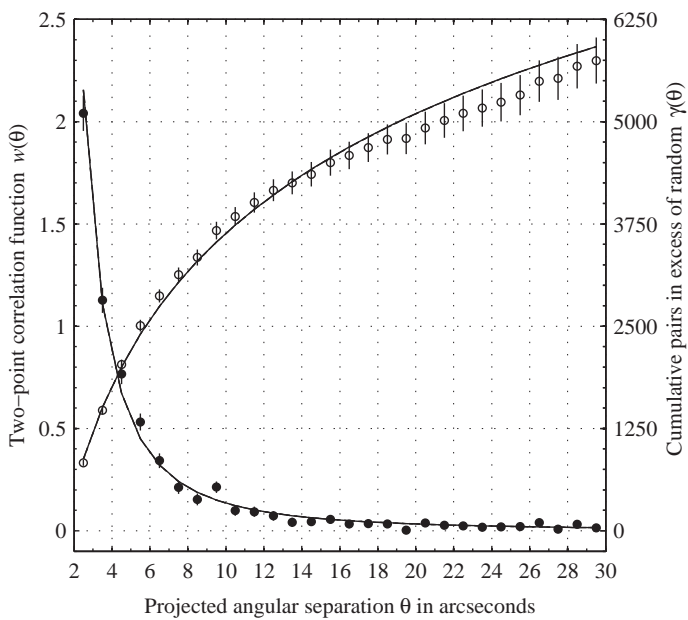}}
      \caption{2PCF as inferred from the total sample (solid circles, left ordinate) and the corresponding CDD (open circles, right ordinate). Poisson errors are indicated as vertical lines. Model curves for the three Galactic structure parameter sets are plotted, too, but as the differences between them are marginal they lie one upon the other, giving a single solid line.}
         \label{allModsTot_b}
   \end{figure}
% Fig. 6,  MCCR_tot_juric_c
   \begin{figure}
   \resizebox{\hsize}{!}{\includegraphics{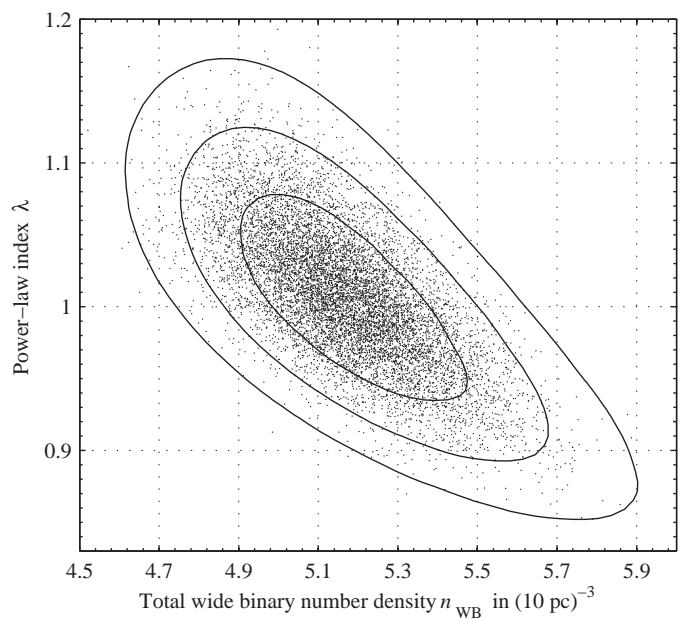}}
      \caption{Distribution of the best-fit values for the total sample using the structure parameter set 2 obtained by the Monte Carlo procedure described in the text. The solid contours (MCCRs) are lines of constant $\chi^2$ and enclose 68.3\%, 95.4\%, and 99.7\% of the best-fit values.}
         \label{MCCR_tot_juric_c}
   \end{figure}
% Fig. 7, MCCR_tot_allMods
   \begin{figure}
   \resizebox{\hsize}{!}{\includegraphics{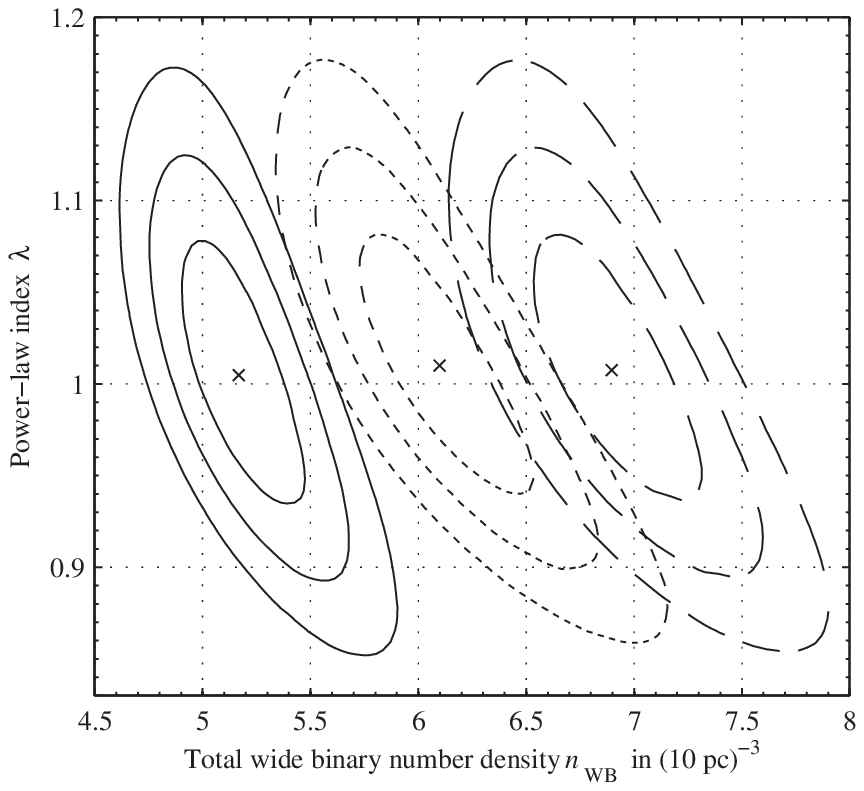}}
      \caption{MCCRs for the three Galactic structure parameter sets corresponding to the total sample. Long dashed line: set 1; solid line: set 2; short dashed line: set 3. The crosses indicate the best-fit values inferred from the true \emph{observed} number of pairs, $F_{\rm corr}(\theta)$.}
         \label{MCCR_tot_allMods}
   \end{figure}
   
% table 2, subsamParam
\begin{table*}
\caption{Parameters of the final sample and of the subsamples}            
\label{subsamParam}      
\centering
\renewcommand{\footnoterule}{}            
\begin{tabular}{cr@{.}lr@{.}llr@{\,}llll}       
\hline\hline                 
(Sub)sample & \multicolumn{2}{l}{$\Omega$} & \multicolumn{2}{l}{$\ell$} & $b$ & \multicolumn{2}{l}{$N_{\rm obs}$} & $D_{\rm eff}$$^{\rm a}$ & $a_{\rm T}$$^{\rm a}$ & $f_{\rm Halo}$$^{\rm a}$  \\
~ & \multicolumn{2}{l}{[deg$^2$]} & \multicolumn{2}{l}{[deg]} & [deg] & \multicolumn{2}{l}{~} & [pc] & [pc] & \%\\ 
\hline                       
total & 668&58 & 175&6 & 81.6 & 669&843 & $1\,555$ & $1.07$ & $30.0$\\
$r<20.0$ mag & 668&58 & 175&6 & 81.6 & 535&595 & $1\,380$ & $1.06$ & $25.2$\\
$r<19.5$ mag & 668&58 & 175&6 & 81.6 & 425&674 & $1\,235$ & $1.06$ & $20.6$\\
left & 334&57 & 191&2 & 73.7 & 326&333 & $1\,495$ & $1.08$ & $28.3$\\
right & 334&01 & 102&5 & 84.8 & 343&510 & $1\,625$ & $1.05$ & $31.0$\\
A & 78&94 & 180&1 & 68.3 & 77&008 & $1\,460$ & $1.09$ & $26.8$\\
B & 78&10 & 167&5 & 75.3 & 76&011 & $1\,510$ & $1.08$ & $28.6$\\
C & 78&99 & 136&0 & 79.8 & 76&686 & $1\,565$ & $1.06$ & $29.9$\\
D & 78&81 & 93&3 & 78.4 & 82&780 & $1\,635$ & $1.05$ & $30.6$\\
E & 88&23 & 208&0 & 69.7 & 87&751 & $1\,485$ & $1.08$ & $27.4$\\
F & 88&30 & 210&6 & 78.6 & 85&563 & $1\,545$ & $1.07$ & $29.5$\\
G & 87&95 & 215&2 & 87.5 & 88&440 & $1\,615$ & $1.05$ & $31.0$\\
H & 88&27 & 33&4 & 83.6 & 95&604 & $1\,700$ & $1.04$ & $31.8$\\
\hline                                   
\end{tabular}
\begin{list}{}{}
\item[$^{\mathrm{a}}$] Calculated using the Galactic structure parameter set 2.
\end{list}
\end{table*}
   
% table 3, bestFitValuesFinal
\begin{table}
\caption{Best-fit values: total sample}            
\label{bestFitValuesFinal}      
\centering
\renewcommand{\footnoterule}{}                  
\begin{tabular}{clllll}       
\hline\hline                 
set of structure & $n_{\rm WB}$ & $\lambda$ & $Q$ & $\chi^2/\nu$ & $f_{\rm WB}$   \cr
parameters & [pc$^{-3}$] & ~ & ~ & ~ & \%\\ 
\hline                       
   set 1 & $0.0052$ & $1.00$ & $0.29$ & $1.14$ & 10.2 \\      
   set 2 & $0.0069$ & $1.01$ & $0.30$ & 1.13 & $13.6$ \\
   set 3 & 0.0061 & $1.01$ & $0.29$ & $1.14$ & $12.1$ \\
\hline                                   
\end{tabular}
\end{table}
 
The results of the analysis for the total sample are shown in Fig.~\ref{allModsTot_b} and Table \ref{bestFitValuesFinal}. Figure \ref{allModsTot_b} shows the 2PCF estimate $\hat{w} _{\rm corr}(\theta)$ and the CDD $\hat\gamma(\theta)$. The statistical uncertainties calculated according to Eq.~\ref{dw} are shown as vertical lines. A strong clustering signal out to at least $\theta=10\arcsec$ is evident, whereas the CDD suggest that there are pairs in excess of a random distribution up to maximum angular separation examined, that is, up to 30\arcsec. 
The outlier at $\theta=9\arcsec$ is probably a random fluctuation.
We also plot in Fig.~\ref{allModsTot_b} best-fitting models using the three Galactic structure parameter sets described in Sect.~\ref{rhoD}.

The best-fit values of the two free parameters, $n_{\rm WB}$ and $\lambda$, are tabulated in Table \ref{bestFitValuesFinal} for the three Galactic structure parameter sets. The power-law index $\lambda$ appears to be quite independent of the set we choose. The wide binary density $n_{\rm WB}$, on the other hand, shows some variation. The difference between the sets 1 and 2 reflects, for the most part, the difference in the overall normalisation $\rho(0)$ of the density distribution, whereas in set 3 the unequal halo normalisation with respect to the other two sets also contributes to the difference in $n_{\rm WB}$.

We also list the goodness-of-fit $Q$ and the corresponding reduced chi-square values ($\chi^2$ divided by the degrees of freedom $\nu$) in Table \ref{bestFitValuesFinal}. All three sets of Galactic structure parameters give equally good fits, whereas the reduced chi-square values, which are only slightly more than unity, indicate that we have not severely underestimated the uncertainty in the 2PCF.

In Fig.~\ref{MCCR_tot_juric_c} we show, representative of the other structure parameter sets, the distribution of the best-fit values from the synthetic samples using set 2, our standard set. 
In the same figure the 68.3\% ($1\sigma$), 95.4\% ($2\sigma$), and 99.7\% ($3\sigma$) MCCRs are also plotted. Quoting 95.4\% confidence intervals throughout, we find for our final sample using the Galactic structure parameter set 2
\begin{equation}
n_{\rm WB}=0.0052_{-0.0005}^{+0.0006}\mbox{ pc}^{-3}\qquad\mbox{and}\qquad\lambda=1.00_{-0.12}^{+0.13}\,.
\end{equation}  
The power-law index $\lambda$ of semi-major axis distribution is consistent with \citeauthor{1924PTarO..25f...1O}'s law ($\lambda=1$) up to the Galactic tidal limit, 
whereas the number density $n_{\rm WB}$ corresponds to a wide binary fraction with respect to \emph{all} stars (\emph{no} colour-cut) in the solar neighbourhood of
\begin{equation}
f_{\rm WB}\equiv\frac{2n_{\rm WB}}{n_*}=10.2_{-1.0}^{+1.2}\,\%,
\end{equation}
where $n_*\simeq0.10$ pc$^{-3}$ is the \emph{total} local stellar number density, i.e.~the integral over the \emph{whole} \citeauthor{1997ESASP.402..675J} LF. (The local wide binary density $n_{\rm WB}$ corresponds only to wide binaries with both components having $g-r>0.5$ mag.)

We show the confidence regions corresponding to the three sets of Galactic structure parameters in Fig.~\ref{MCCR_tot_allMods}. 
The confidence intervals for $\lambda$ agree for all the sets,
which demonstrates that we can determine $\lambda$ and its confidence intervals reliably -- provided that no systematic error has crept into our analysis. 
The wide binary density $n_{\rm WB}$, on the other hand, is more sensitive to the exact values of the Galactic structure parameters than $\lambda$ is, so the values we have derived for the wide binary density $n_{\rm WB}$ and the fraction $f_{\rm WB}$ should be viewed with caution. We are, however, confident that the true value of the wide binary density $n_{\rm WB}$ is within a factor of 2 of our derived value.

Since the structure parameter set 2 is -- as far as known to the authors -- the only one systematically corrected for unresolved multiplicity, we give that set more weight. From now on, we use the bias-corrected values from \citet{2008ApJ...673..864J}, i.e.~our standard set 2, exclusively.

% Fig. 8, psi_s
 \begin{figure}	
   \resizebox{\hsize}{!}{\includegraphics{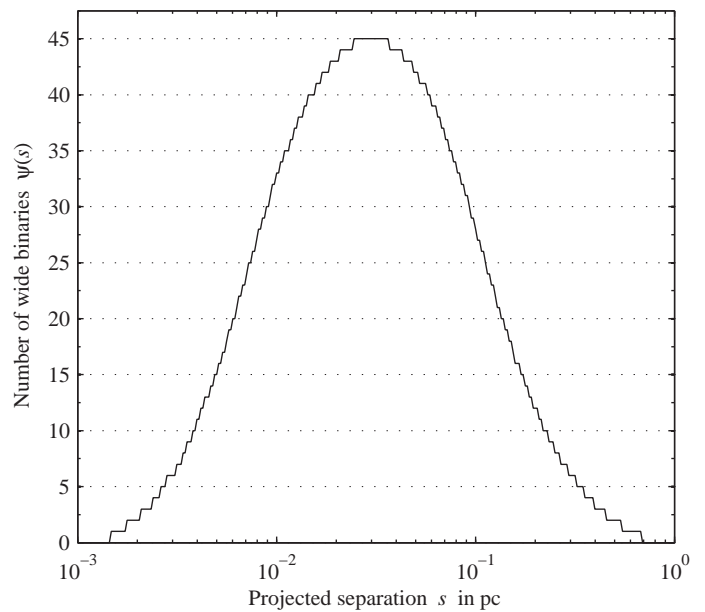}}
      \caption{Distribution of wide binaries as a function of projected separation $s$ that is expected to be observed in the total sample, given the model assumptions and the selection criteria we have adopted (see text).}
         \label{psi_s}
   \end{figure}
Having determined the two free parameter in our model, we can calculate the number of wide binaries in our total sample as a function of the projected separation $s$. We show the distribution in Fig.~\ref{psi_s} whose shape is largely dominated by the effective volume, i.e.~by selection effects. 
In particular, this distribution implies that we have observed $830^{+250}_{-215}$ very wide binaries with a projected separation larger than 0.1 pc in our total sample, while none are expected to be found beyond 0.8 pc, given our selection criteria. However, that extremely wide binaries with projected separations of more than 1 pc can exist in the Galactic halo has recently been confirmed by \citet{2009MNRAS.396L..11Q}. 

\subsection{Differentiation in terms of apparent magnitude}
% Fig. 9, 195_200
 \begin{figure}	
   \resizebox{\hsize}{!}{\includegraphics{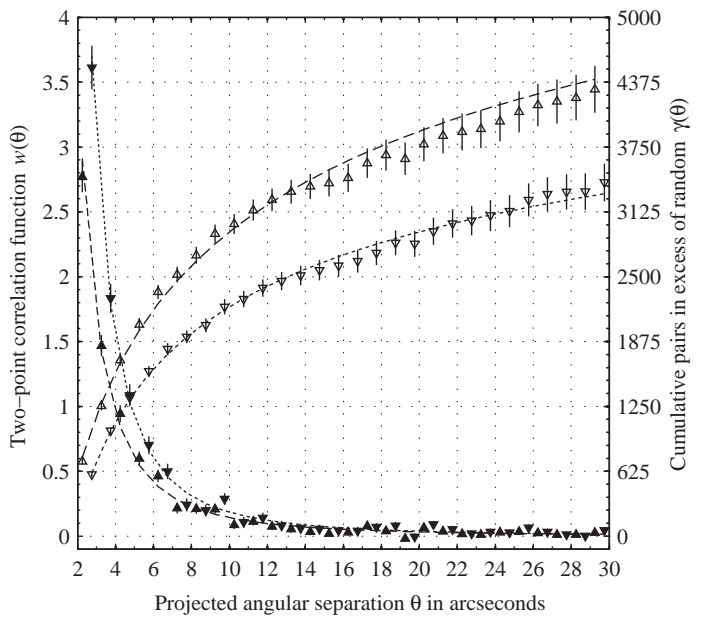}}
      \caption{2PCF and CDD for the $r<20.0$ mag and $r<19.5$ mag subsamples together with the corresponding model curves. Up-pointing triangles (long-dashed lines) are for the $r<20.0$ mag subsample, whereas down-pointing triangles (short-dashed lines) are for the $r<19.5$ mag subsample. The symbols were shifted apart by $0.25\arcsec$ for better visibility.}
         \label{195_200}
   \end{figure}
% Fig. 10,  MCCR_195_200
   \begin{figure}
   \resizebox{\hsize}{!}{\includegraphics{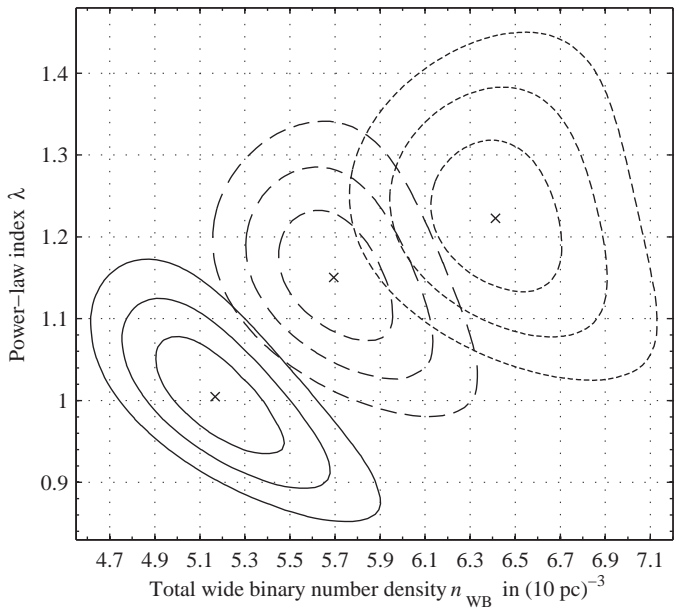}}
      \caption{MCCRs for the $r<20.0$ mag (long dashed contours) and $r<19.5$ mag (short-dashed contours) subsamples. For comparison, the MCCR for our final sample is also shown (solid-contours). The crosses have the same meaning as in Fig.~\ref{MCCR_tot_allMods}.}
         \label{MCCR_195_200}
   \end{figure}
   
   % table 3, bestFitValuesMag
\begin{table}
\caption{Best-fit values: $r<20.0$ mag and $r<19.5$ mag}            
\label{bestFitValuesMag}      
\centering
\renewcommand{\footnoterule}{}             
\begin{tabular}{clllll}       
\hline\hline                 
Subsample & $n_{\rm WB}$ & $\lambda$ & $Q$ & $\chi^2/\nu$ & $f_{\rm WB}$   \cr
~ & [pc$^{-3}$] & ~ & ~ & ~ & \%\\ 
\hline                        
$r<20.0$ mag & 0.0057 & $1.15$ & $0.18$ & $1.25$ & $11.3$ \\
$r<19.5$ mag & $0.0064$ & $1.22$ & $0.34$ & $1.09$ & 12.7 \\ 
\hline                                   
\end{tabular}
\end{table}

To test the consistency of the standard model, we used two subsamples having brighter upper apparent magnitude limits than our total sample and analysed them in the same way as the total sample itself. We set the upper apparent magnitude limit to $r=20.0$ mag and $r=19.5$ mag -- all other selection criteria (including the lower apparent magnitude limits) remain unchanged. The main parameters of these subsamples are listed in Table \ref{subsamParam}.

In Fig.~\ref{195_200} we show the 2PCF and the CCD as in Fig.~\ref{allModsTot_b}, together with the corresponding model curves for standard set 2 (see caption). 
The best-fit values are listed in Table \ref{bestFitValuesMag}.
The MCCRs of the $r<20.0$ mag and $r<19.5$ mag subsamples are shown in Fig.~\ref{MCCR_195_200}.
 
If the model were self-consistent, we would expect that the best-fit values agree with each other within their uncertainties. Figure~\ref{MCCR_195_200} indicates that this is not the case: It appears that the best-fit values are systematically shifted to higher densities and larger power-law indices when using a brighter upper magnitude limit. As we discuss in Sect.~\ref{incons}, this inconsistency is most likely an artefact of the (oversimplifying) model assumptions and does not entirely undermine our results.
\subsection{Differentiation in terms of direction}\label{dirrDiff}
% Fig. 11,  leftRight_2b
   \begin{figure}
   \resizebox{\hsize}{!}{\includegraphics{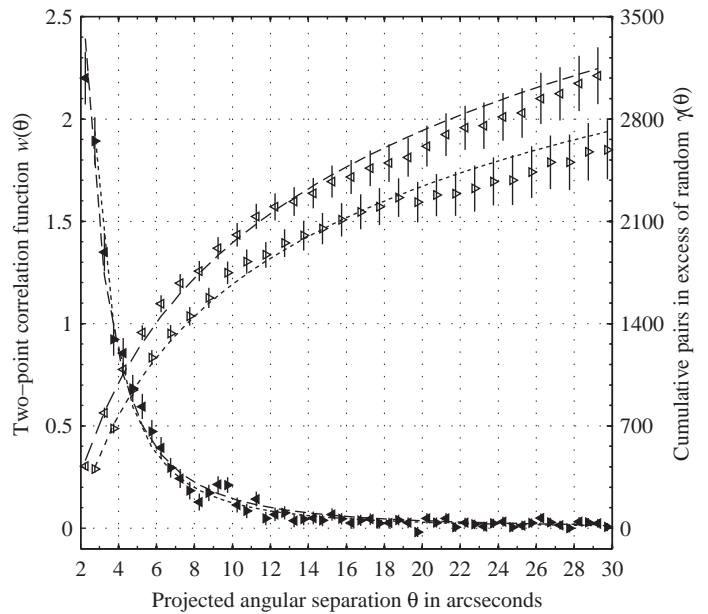}}
      \caption{2PCF and CDD for the left and the right subsamples, together with the corresponding model curves. Left-pointing triangles (long-dashed lines) are for the left subsample, whereas right-pointing triangles (short-dashed lines) are for the right subsample. The symbols were shifted apart by $0.25\arcsec$ for better visibility.}
         \label{leftRight_2b}
   \end{figure}
% Fig. 12,  MCCR_LeftRight
   \begin{figure}
   \resizebox{\hsize}{!}{\includegraphics{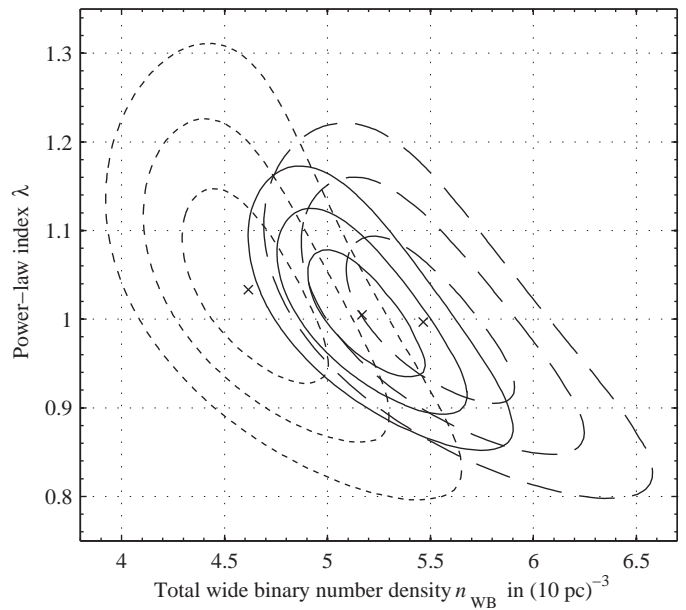}}
      \caption{MCCRs for the left (long-dashed contours) and right (short-dashed contours) subsamples. For comparison, also the MCCR for our final sample is shown (solid contours). The crosses have the same meaning as in Fig.~\ref{MCCR_tot_allMods}.}
         \label{MCCR_LeftRight}
   \end{figure}

% table 4,  bestFitValuesLR
\begin{table}
\caption{Best-fit values: left and right}            
\label{bestFitValuesLR}      
\centering
\renewcommand{\footnoterule}{}              
\begin{tabular}{clllll}       
\hline\hline                 
Subsample & $n_{\rm WB}$ & $\lambda$ & $Q$ & $\chi^2/\nu$ & $f_{\rm WB}$   \cr
~ & [pc$^{-3}$] & ~ & ~ & ~ & \%\\ 
\hline                       
left & 0.0055 & $1.00$ & $0.55$ & $0.94$ & 10.8 \\
right & $0.0046$ & $1.03$ & $0.82$ & $0.74$ & \ \ 9.1 \\      
\hline                                   
\end{tabular}
\end{table}

In a previous study, \citet{1989MNRAS.237..311S} found that the binaries appear to be highly clumped in the NGP. 
However, it has not become entirely clear whether this patchiness of the wide binary distribution in the sky is a real physical characteristic of the wide binary population or if it is due to statistical fluctuations. 
In principle, we can check whether the wide binary density varies  with position in the sky by dividing our sample into sub\-areas.

To begin with, we divide our sample in two halves by cutting it along the $\alpha=185^{\circ}$ meridian. In the following, we refer to the subarea with $\alpha<185^{\circ}$ as the ``left'' subsample and the other half as the ``right'' subsample. The subsamples' main parameters are listed in Table \ref{subsamParam}. 

In Fig.~\ref{leftRight_2b} we show the 2PCFs inferred from the left and the right sample, as well as the corresponding CDDs. 
It appears that there are a few more pairs in excess of random in the left half.
In the same figure also the corresponding model curves are plotted. The best-fit values are listed in Table \ref{bestFitValuesLR}.
 
In Fig.~\ref{MCCR_LeftRight} we show the MCCRs of the left and the right subsamples. 
The left subsample indeed shows a higher wide binary density than the right one. The difference is significant at the $3\sigma$ level. 
The power-law indices, on the other hand, do agree in both subsamples.
Regarding the total sample, the right half differs significantly (at $3\sigma$), whereas the left half is more consistent with it.

\begin{figure*}
 \centering
   \resizebox{\hsize}{!}{\includegraphics{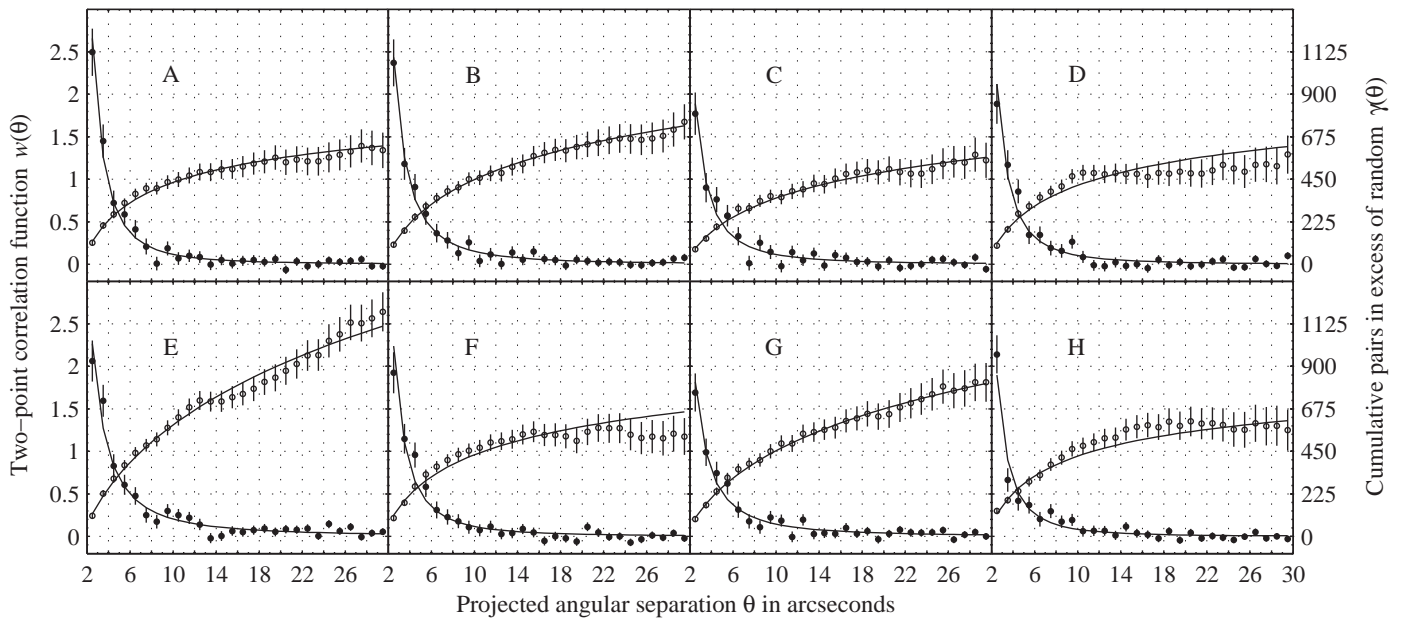}}
      \caption{2PCF (solid circles) and CDD (open circles) for the subsamples A to H, together with the corresponding model curves. The subfigures are arranged as the corresponding subareas would be seen on the sky.}
         \label{AtoH_2}
 \end{figure*}

The difference in the wide binary density between the left and the right halves is probably a real feature, as any inadequacies of our model (e.g.~inaccuracies of the stellar density distribution we use) should affect both halves in almost the same manner. 
To examine this apparent positional dependency of the wide binary density in more detail, we divide our sample further into eight subsamples, each covering $10^{\circ}\times10^{\circ}$. They are labelled from A (``upper left'') to H (``lower right'') as suggested in Fig.~\ref{AtoH_2}, where the abscissa can thought of representing the right ascension from $165^{\circ}$ (``left'') to $205^{\circ}$ (``right'') and similarly the ordinate represents the declination from $22^{\circ}$ (``bottom'') to $42^{\circ}$ (``top''). The subsample's main parameters are again summarised in Table \ref{subsamParam}.

% Fig. 13,  AtoH_2

% table 5,  bestFitValuesAtoH
\begin{table}
\caption{Best-fit values: A to H}            
\label{bestFitValuesAtoH}      
\centering
\renewcommand{\footnoterule}{}               
\begin{tabular}{cllllr@{.}l}       
\hline\hline                 
Subsample & $n_{\rm WB}$ & $\lambda$ & $Q$ & $\chi^2/\nu$ & \multicolumn{2}{l}{$f_{\rm WB}$}   \cr
~ & [pc$^{-3}$] & ~ & ~ & ~ & \multicolumn{2}{l}{\%}\\ 
\hline                       
A & 0.0050 & $1.27$ & $0.92$ & $0.64$ & 9&9 \\
B & $0.0054$ & $1.00$ & $0.96$ & $0.57$ & 10&7 \\ 
C & $0.0041$ & $1.02$ & $0.36$ & $1.07$ & 8&2\\
D & 0.0046 & $1.19$ & $0.29$ & $1.13$ & 9&1\\
E & 0.0077 & $0.73$ & $0.24$ & $1.18$ & 15&2 \\
F & 0.0045 & $1.14$ & $0.37$ & $1.07$ & 8&9\\
G & $0.0054$ & $0.83$ & $0.67$ & $0.86$ & 10&6 \\
H & $0.0042$ & $1.27$ & $0.64$ & $0.88$ & 8&4 \\      
\hline                                   
\end{tabular}
\end{table}

Figure \ref{AtoH_2} shows the 2PCFs estimate and the CDDs with the best-fit model curves. The corresponding best-fit parameters are listed in Table \ref{bestFitValuesAtoH}. Some subtle differences are apparent between different subsamples. While the CDD in A, B, C, and G are reproduced well by the model, the CDD in D, F, and H appears to be too flat. In those subsamples it seems that almost no physical pairs are present at angular separations over $15\arcsec$, in agreement with the findings of \citet{2008ApJ...689.1244S} (see Sect.~\ref{excons}). 

The subsample E appears to be an outlier, because it contains almost twice as many pairs in excess of random as the seven other subsamples. The listed best-fit values confirm the peculiar character of subsample E having the highest wide binary density of all eight subsamples, together with the lowest power-law index. 
As far as the authors can judge, there is no obvious special feature (e.g.~open star cluster) in subsample E that could cause this anomaly. 

   % Fig. 14,  MCCR_AtoH
   \begin{figure}
   \resizebox{\hsize}{!}{\includegraphics{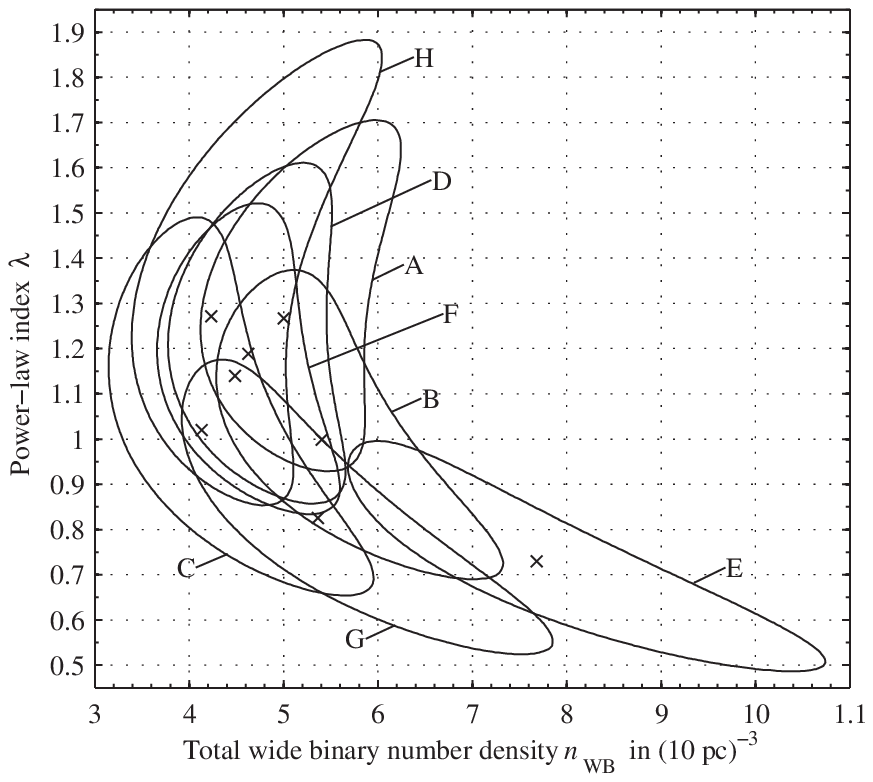}}
      \caption{95.4\% MCCRs for the subsamples A to H. The crosses have the same meaning as in Fig.~\ref{MCCR_tot_allMods}.}
         \label{MCCR_AtoH}
   \end{figure}
   
The $2\sigma$ MCCR are shown in Fig.~\ref{MCCR_AtoH}. Except for the outlier E, all the other subsamples are quite consistent with each other. Also, no obvious trend, e.g.~with Galactic latitude $b$, is apparent. To what extent is the subsample E responsible for the difference between the right and the left subsamples?

% table 6,  bestFitValuesExE
\begin{table}
\caption{Best-fit values: excluding subsample E}            
\label{bestFitValuesExE}      
\centering
\renewcommand{\footnoterule}{}             
\begin{tabular}{clllll}       
\hline\hline                 
Subsample & $n_{\rm WB}$ & $\lambda$ & $Q$ & $\chi^2/\nu$ & $f_{\rm WB}$   \cr
~ & [pc$^{-3}$] & ~ & ~ & ~ & \%\\ 
\hline                       
total\textbackslash E & $0.0049$ & $1.06$ & 0.40 & $1.04$ & $9.6$ \\
left\textbackslash E & $0.0049$ & $1.12$ & $0.76$ & $0.79$ & 9.8 \\     
\hline                                   
\end{tabular}
\end{table}

To answer this question, we repeated the analysis of the total and left (sub)sample excluding subsample E. The results are listed in Table \ref{bestFitValuesExE}. The wide binary density drops significantly to a value more consistent with the right subsample. 
Thus, we conclude that the difference between the right and the left subsamples is largely caused by the outlier E. 
Apart from subsample E, the wide binary densities in different directions appear to be consistent with each other.
The reason for the high density in subsample E remains unclear. However, a statistical fluctuation cannot be ruled out to a level better than $2\sigma$.

%__________________________________________________________________

\section{Discussion}\label{discuss}
% Fig. 15, CF_BPL
   \begin{figure}
   \resizebox{\hsize}{!}{\includegraphics{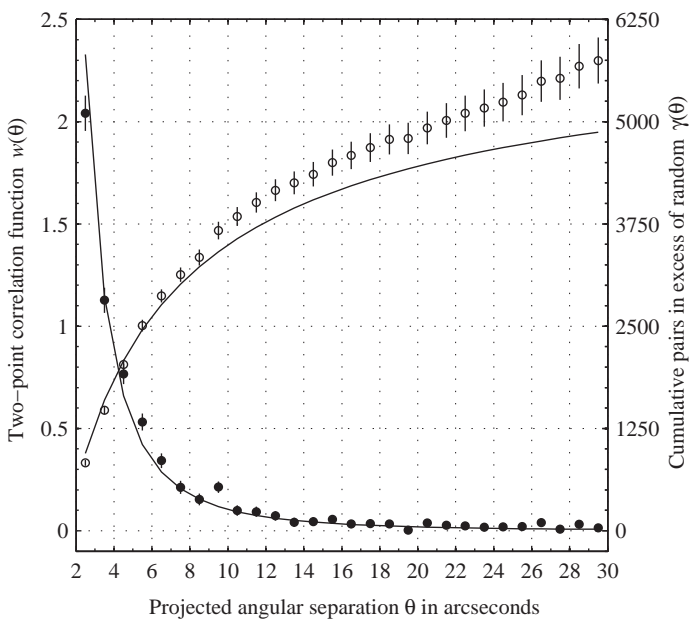}}
      \caption{2PCF and CDD as inferred from our total sample as in Fig.~\ref{allModsTot_b}. The model curve is for the broken power-law model from  \citet{2007AJ....133..889L} (see text).}
         \label{CF_BPL}
   \end{figure}
\subsection{External (in)consistencies}\label{excons}
In this section we compare our results with those of previous studies. Our first point concerns the {\em observed} angular 2PCF. From the CDD of our total sample shown in Fig.~\ref{allModsTot_b} we have noted pairs in excess of random up to $\theta_{\max}=30\arcsec$. This is contradictory to \citet{2008ApJ...689.1244S} who find that there are essentially no physically bound pairs with an angular separation $\theta>15\arcsec$. (We note, however, that this is in accord with some of our \emph{sub}samples, see Fig.~\ref{AtoH_2}.)
We considered several possible reasons for this apparent discrepancy, such as differences in the selection criteria of the \citeauthor{2008ApJ...689.1244S} sample with respect to our sample, underestimation of the total area of holes in our sample, or an overcorrection of edge effects due to those holes, but we convinced ourselves that none of them can account for this disagreement.
A real physical difference in the wide binary population studied, e.g.~caused by the fact that \citeauthor{2008ApJ...689.1244S}'s sample is largely disc-dominated, while our final sample has a substantial halo contribution of $\sim30\%$, can be excluded, as previous studies \citep{2002AJ....124.1144L,2004ApJ...601..289C} found that the disc and halo wide binary populations are reasonably consistent in their statistical characteristics.

Regarding the wide binary fraction, \citeauthor{2008ApJ...689.1244S} also find a much smaller wide binary fraction of below 1\% (decreasing with height above the Galactic plane), as compared to our $f_{\rm WB}\approx$ 10\%. Even if all pairs with a semi-major axis larger than 3\,000 AU (beyond the ``break'') were ignored, we would still find a wide binary fraction of 4.6\%. 
On the other hand, \citet{2007AJ....133..889L} give a binary fraction of at least 9.5\% for separations larger than 1\,000 AU, which is in rough agreement with our results (setting $a_{\min}=1\,000$ AU we find $f_{\rm WB}\approx$ 8.3\%). 

Studying a much smaller region containing brighter stars as compared to our sample, \citetalias{1987ApJ...312..390W} and \citet{1988ApJ...335L..47G,1991PhDT.........8G} found an unphysically large wide binary density in the direction of the NGP. \citet{1989MNRAS.237..311S} attribute this overdensity to a large statistical fluctuation. The region where this overdensity was found would be located in our subarea H. But as we probe fainter stars, no density enhancement is evident in that subarea. The slight overdensity we found in subarea E is different from that previously noted by \citetalias{1987ApJ...312..390W} and \citeauthor{1991PhDT.........8G}. 

As to the separation distribution, previous studies have found that the observational data are described by \citeauthor{1924PTarO..25f...1O}'s law ($\lambda=1$) up to a certain maximum separation (``break''): \citet{2007AJ....133..889L} find this break to be around 3\,500 AU, beyond which a steeper slope, with $\lambda\approx1.6$, should apply. \citet{2008ApJ...689.1244S} basically agree with \citeauthor{2007AJ....133..889L} and find in addition that the maximum separation increases with height above the Galactic plane. \citet{2004RMxAC..21...49P} divided their wide binary catalogue \citep{1994RMxAA..28...43P} into two subsamples consisting of the oldest and youngest systems, respectively, and find that the ``maximum major semiaxis for which \citeauthor{1924PTarO..25f...1O}'s distribution holds is much larger for the youngest binaries ($a_m=7862$ AU) than for the oldest ($a_m=2409$ AU)''.
\citet{2004ApJ...601..289C} find the data to be described by a single powerlaw with an index of approximately 1.6 for their disc and halo samples.
However, they note a ``puzzling'' flattening of the distribution of the disc binaries between 10\arcsec and 25\arcsec. As already suggested by \citeauthor{2008ApJ...689.1244S}, this flat range might be the domain where \citeauthor{1924PTarO..25f...1O}'s law is valid. 
In view of the substantial uncertainties inherent in studies by \citeauthor{2007AJ....133..889L} and by \citeauthor{2004ApJ...601..289C}, the two studies appear to be broadly consistent.
Similar to our study, \citet{1991PhDT.........8G} assumed that the semi-major axis distribution is described by a single powerlaw. He finds the power-law index to be $0.7\pm0.2$ for his NGP sample covering nearly 240 square degree. At intermediate Galactic latitudes he finds the slope to be steeper: $\lambda=1.3\pm0.2$. His data seem to favour a (somewhat unrealistic\footnote{``Physically, one doesn't expect to see a sharp cutoff.'' \citep{1990bdm..proc..117W}}) ``cutoff'' around 0.1 pc, especially for the NGP sample.

The question of a break in the separation distribution has much been discussed in the context of DM constraints. \citet{1991ApJ...382..149W}, for example, show that the observational data suggest a break but they do not \emph{require} one statistically. 
Looking at the CDD for our total sample in Fig.~\ref{allModsTot_b}, we note that our best-fit model slightly overestimates the number of pairs in excess of random at larger angular separations, and a more curved model line would be preferred by the data. This could indeed be interpreted as a hint that the semi-major axis distribution is broken, because having a steeper power-law index from a certain semi-major axis on would result in a flatter CDD model curve. In principle, we could easily fit a broken powerlaw to our data as well. However, too many free parameters will only destabilise our modelling.   
Instead we check whether our data are also consistent with the specific broken power-law distribution found by \citet{2007AJ....133..889L}. 
To this end we use a broken reduced semi-major axis distribution of the form
\begin{equation}
q(a)=q_1(a)+q_2(a)=c_{\lambda_1}a^{-\lambda_1}\Theta(a_{\rm c}-a)+c_{\lambda_2}a^{-\lambda_2}\Theta(a-a_{\rm c})
\end{equation}
with the Heaviside function $\Theta$ that introduces a break at semi-major axis $a_{\rm c}$.
It is normalised, as in \S\ref{redSepDis}, according to
\begin{equation}
\int_{a_{\min}}^{a_{\rm T}}q(a)\,\mbox{d}a=\int_{a_{\min}}^{a_{\rm c}}q_1(a)\,\mbox{d}a+\int_{a_{\rm c}}^{a_{\rm T}}q_2(a)\,\mbox{d}a=1\,.
\end{equation}
Using the values given by \citet{2007AJ....133..889L}, $\lambda_1=1$, $\lambda_2=1.6$, and $a_{\rm c}=0.02$ pc, we find for the normalisation constants
\begin{equation}
c_{\lambda_1}=\left[\ln a_{\rm c}-\ln a_{\min}+\frac{\left(a_{\rm T}/a_{\rm c}\right)^{1-\lambda_2}-1}{1-\lambda_2}\right]^{-1}
\end{equation}
and
\begin{equation}
c_{\lambda_2}=c_{\lambda_1}a_{\rm c}^{\lambda_2-1}\,.
\end{equation}
The distribution of the projected separations $Q(s)$ then splits into a sum, too:
\begin{equation}
Q(s)=\left(\frac{s}{\mbox{pc}}\right)^{-\lambda_1}C_{\lambda_1}(s)\,~\mbox{pc}^{-1}+\left(\frac{s}{\mbox{pc}}\right)^{-\lambda_2}C_{\lambda_2}(s)\,~\mbox{pc}^{-1},
\end{equation}
where $C_{\lambda_i}(s)$ is defined as in Eq.~\ref{Cs} with the limits chosen appropriately when integrating over $\eta$.

We now keep the parameter given by \citeauthor{2007AJ....133..889L} fixed so that our broken power-law model has only one free parameter: the wide binary density $n_{\rm WB}$.
In Fig.~\ref{CF_BPL} we show the best-fit model curves corresponding to $n_{\rm WB}=0.0075$ pc$^{-3}$. The broken power-law model also gives a decent fit to the 2PCF inferred from our total sample. (One should not be misled by large discrepancy between the model and the CDD showing the \emph{cumulative} difference between the 2PCF and the corresponding model curve.) With a goodness-of-fit of $Q\simeq0.0047$ and a reduced chi-square of $\chi^2/\nu\simeq1.85$, the broken power-law model provides a significantly worse bestfit to the data than our single power-law model. 
Nevertheless, as the errors in the 2PCF are slightly underestimated, the model is not put into question until $Q<0.001$ \citep[][their \S15]{1992nrca.book.....P}, and we conclude that the broken power-law distribution found by \citet{2007AJ....133..889L} can not be rejected with confidence either.

\subsection{Internal (in)consistencies}\label{incons}
Figure \ref{MCCR_195_200} shows that the best-fitting values for $\lambda$ and $n_{\rm WB}$ are inconsistent, within the adopted {\em random} errors, when the limiting magnitude is varied. Both parameters are systematically shifted by roughly 20\% when the limiting magnitude is changed by 1 mag. Further inconsistencies were met when we tried to differentiate our total sample with respect to colour (not shown here). This very likely means that there are unaccounted for {\em systematic} errors in the modelling, resting on oversimplified assumptions. 

A basic assumption on the properties of wide binaries was that both components are drawn at random from the same stellar LF. There is, however, some evidence that this is not quite correct: 
\citet{1995ApJ...441..200G} noticed that ``binaries are bluer and more distant than one would expect if they were formed of random combinations of field stars''.
Also \citet{2007AJ....133..889L} find ``that the luminosity function of the secondaries is significantly different from that of the single stars field population, showing a relative deficiency in low-luminosity ($8<M_V<14$) objects''. Similarly, \citet{2008ApJ...689.1244S} report that blue stars ($g-i\la2.0$ mag, corresponding roughly to $g-r\la1.4$ mag) that are a member of a wide binary, have more blue companions than expected from the LF. (For red stars, however, they find that the components are drawn randomly from the LF.)
Moreover, it has long been known \citep[e.g.][their Fig.~2]{1980ApJS...44...73B} that stars of early spectral type have smaller scale heights than late type stars; therefore, the assumption that the stellar density distribution depends only on distance, and not on the luminosity of the stars, might be an oversimplification as well.  

On the other hand, the internal inconsistency discussed is overemphasised in Fig.~\ref{MCCR_195_200} because we only include the statistical errors stemming from the pair counts when determining the MCCRs. If we also included the errors in the Galactic structure parameters and those from the LF, the MCCRs would be considerably larger, and the inconsistency evident from Fig.~\ref{MCCR_195_200} would not be as severe as it seems. That a variation in the Galactic structure parameter has a significant impact at least on $n_{\rm WB}$, is apparent from Fig.~\ref{MCCR_tot_allMods}. The systematic drift towards larger $\lambda$ and $n_{\rm WB}$, when adopting a brighter upper apparent magnitude limit (cf.~Fig.~\ref{MCCR_195_200}), still points to some inconsistencies in our model.
We think, however, that these inconsistencies do not entirely undermine our major results, although the quoted uncertainties might be considerably larger.

\section{Summary and conclusions}\label{conclude}
We have derived the angular 2PCF for nearly 670\,000 SDSS stars brighter than $r$ = 20.5 mag and redder than $g-r$ = 0.5 mag in a region of approximately 670 square degrees around the NGP. Various corrections had to be made for quasar contamination, survey holes, and bright stars. There is an unambiguous correlation signal on small scales up to 30\arcsec. We modelled this signal by a modified WW-technique, closely following a previous study of \citet{1988ApJ...335L..47G,1991PhDT.........8G} that was based on a much smaller sample of stars. 
The modelling involved a number of parametrised distribution functions: the spatial density and LF of stars, as well as the separation distribution of binaries. The Galactic model used is based on the recent study by \citet{2008ApJ...673..864J} and the stellar LF derived by \citet{1997ESASP.402..675J}. For the wide binary semi-major axis distribution we assumed a single powerlaw. Furthermore, essential assumptions are: (1) binary stars follow the same density distribution as single stars, (2) both components of a binary are randomly drawn from the single star luminosiy function. These assumptions allowed a significant simplification of the model. The Galactic structure parameters were fixed (we explore three different sets), while the local wide binary number density $n_{\rm WB}$ and the power-law index $\lambda$ of the semi-major axis distribution were left free; i.e., they have been determined by a least-squares fitting algorithm.

The best fit to the observed angular 2PCF of the total sample was obtained with $\lambda\approx1.0$, which corresponds to the canonical \citeauthor{1924PTarO..25f...1O} law, and $n_{\rm WB}\approx0.005$ pc$^{-3}$, meaning an overall local wide binary fraction of about 10.0\% in the projected separation range of 0.001 pc (200 AU) to 1 pc (Galactic tidal limit). Previous studies \citep{2004RMxAC..21...49P,2007AJ....133..889L, 2008ApJ...689.1244S} have also found the data to be consistent with \citeauthor{1924PTarO..25f...1O}'s law, but only to a maximal separation that is considerably smaller than the Galactic tidal limit. Beyond that maximum separation, the distribution continues with a steeper decline. We have shown that our data are also consistent with the broken powerlaw found by \citeauthor{2007AJ....133..889L}, although with a fit of lower quality (though involving only one free parameter, namely, $n_{\rm WB}$). Given this ambiguity, we are not able to put any strong constraints on the presence of a break in the wide binary separation distribution, which is regarded as one of the most interesting aspects of wide binaries. As to the wide binary fraction, we are in good accord with \citet{2007AJ....133..889L}, whereas an apparent discrepancy with \citet{2008ApJ...689.1244S} remains.

A differentiation of the sample with respect to limiting apparent magnitude turned up a systematic dependence of the binary parameters on the sample depth, which is most probably an artefact caused by oversimplified model assumptions. This conjecture is strengthened by the impossibility obtaining self-consistent results when differentiating with respect to colour. We conclude that one or more of the simplifying assumptions put into the model (e.g. that the luminosities of the binary components are independent of each other and draw randomly from the single-star LF or that binaries' density distribution follows exactly the single-star density distribution and that the density distribution is independent of the stars' luminosities) are not quite correct.

Differentiating the sample in terms of direction in the sky did yield some modest but non-significant variations. Only in one direction (subarea E) was an unexplained overdensity found.

While we have shown here that the stellar angular 2PCF, as a complement to common proper motion studies, basically works and remains a viable tool for the study of wide binaries, it has become clear that this method is severely limited by the need for -- even more -- complex modelling.
To relax those simple model assumptions would unduly complicate the analysis much further and probably no longer yield unique solutions. Any more efficient progress will indeed have to involve distance information to discriminate against unwanted chance projections. We plan to include distance information in our future work. 
In spite of the limitations of the present, simple modelling of the angular 2PCF, we think that the general result for the total sample derived here, i.e. $\lambda \approx1$ (\citeauthor{1924PTarO..25f...1O} law) and $f_{\rm WB} \approx10\%$ among stars having a spectral type later than G5, to within an uncertainty of 10\%-20\%, is a robust result.

\appendix
\section{Edge correction for holes}\label{EC4H}
% Fig. A.1, holeCorr
   \begin{figure}
   \resizebox{\hsize}{!}{\includegraphics{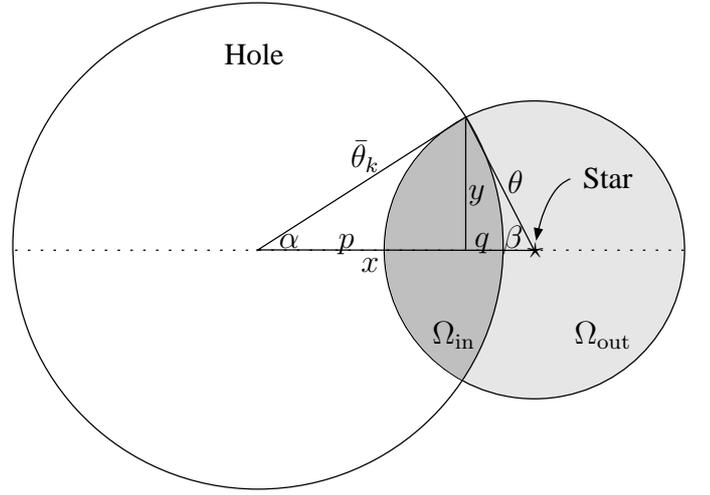}}
      \caption{Illustration of the geometry used to correct for edge effects due to holes. The large circle represents the bounding circle of a hole or a bright star mask. A star lying close to the hole is indicated. The region within a distance $\theta$ of that star (highlighted in grey) is partially inside the hole (dark grey intersection).}
         \label{holeCorr}
   \end{figure}
To quantify the edge effect from holes in our sample, we assume that the holes are circular and flat (Euclidian) and that effects due to intersecting holes are negligible. 
Let $\overline{A}_k(\theta)$ be the solid angle of the annulus of width $2\theta$ around a hole of radius $\bar{\theta}_k$
\begin{equation}
\overline{A}_k(\theta)=4\pi\theta\left(\bar{\theta}_k+\theta\right)\,,\qquad\mbox{with}\quad k=\{\rm{SH,BS}\}\,,
\end{equation}
where $k$ stands for hole masks (survey holes: SH) or bright star masks (BS). Due to the holes in our sample we observe only a fraction $F_{\rm H}^{\rm tot}(\theta)$ of all stellar pairs in the residual solid angle $\Omega$ sparated by an angular distance $\theta$
\begin{equation}
F_{\rm H}^{\rm tot}(\theta)=1-\frac{1}{\Omega}\sum_{k=\{\rm{SH,BS}\}}N_k\overline{A}_k(\theta)\left(1-F_{\rm H}(\theta; \bar{\theta}_k)\right)\,,
\end{equation}
where $F_{\rm H}(\theta; \bar{\theta}_k)$ is the fraction of all pairs separated by $\theta$ we observe in $\overline{A}_k(\theta)$
\begin{equation}\label{FH}
F_{\rm H}(\theta; \bar{\theta}_k)=\frac{2\pi}{\overline{A}_k(\theta)}\int_{\bar{\theta}_k}^{\bar{\theta}_k+\theta}xf(x;\theta,\bar{\theta}_k)\mbox{d}x\,.
\end{equation}
Here, $f$ is the fraction of the area $\pi\theta^2$ of the disc with radius $\theta$ around a star that lies outside the hole (Fig.~\ref{holeCorr})
\begin{equation}
f(x;\theta,\bar{\theta}_k)=\frac{\Omega_{\rm out}(x;\theta,\bar{\theta}_k)}{\pi\theta^2}=1-\frac{\Omega_{\rm in}(x;\theta,\bar{\theta}_k)}{\pi\theta^2}\,,
\end{equation}
whereas $\Omega_{\rm in}=\pi\theta^2-\Omega_{\rm out}$ is the area of that disc within the hole, i.e.~the intersection. With the nomenclature given in Fig.~\ref{holeCorr}, we find
\begin{equation}
\Omega_{\rm out}=\bar\theta_k^2\alpha-yp+\theta^2\beta-yq\,,
\end{equation}
where
\begin{equation}
\alpha=\arcsin\left(\frac{y}{\bar\theta_k}\right)\,;\quad\beta=\arcsin\left(\frac{y}{\theta}\right)
\end{equation}
and
\begin{equation}
p=\sqrt{\bar\theta_k^2-y^2}\,;\quad q=\sqrt{\theta^2-y^2}\,.
\end{equation}
Solving $x=p+q$ for $y$ using \emph{Maple}  
gives
\begin{equation}
y=\frac{1}{2x}\sqrt{2\theta_k^2x^2-x^4-\theta^4+2\theta^2x^2+2\theta^2\theta_k^2-\theta_k^4}\,,
\end{equation}
where the integration in (\ref{FH}) was performed with \emph{Maple}, too.

\section{Galactic tidal limit}\label{aT}
A crude estimate of the Galactic tidal limit $a_{\rm T}$, i.e.~of the maximum semi-major axis $a_{\max}$ of a binary star with total mass ${\mathsf M}$ orbiting in the Galactic tidal field is provided by the \emph{Jacobi limit}
$r_{\rm J}$ (sometimes also referred to as \emph{zero-velocity surface} or \emph{Roche sphere}) \citep[e.g.][\S8.3]{2008gady.book.....B}
\begin{equation}
r_{\rm J}\simeq\left(\frac{{\mathsf M}}{3{\mathcal M}(<\!\widetilde{D})}\right)^{\frac{1}{3}}\widetilde{D}\,,
\end{equation}
where $\widetilde{D}$ is the galactocentric distance of the binary system and ${\mathcal M}(<\!\widetilde{D})$ is the Galactic mass enclosed within radius $\widetilde{D}$. Then, ${\mathcal M}(<\!\widetilde{D})$ is given by
\begin{equation}
{\mathcal M}(<\!\widetilde{D})=\frac{v_c^2}{G}\widetilde{D}\simeq 1.1\cdot 10^7\left(\frac{\widetilde{D}}{\mbox{pc}}\right){\mathcal M}_{\odot}\,,
\end{equation}
where we have adopted the canonical value of the circular speed $v_c=220$ kms$^{-1}$.

For a rough estimate of this distance range, it is useful to define the \emph{effective depth} of the sample as the most likely distance of an arbitrarly chosen star \citep{1988ApJ...329..253W}. 
In a magnitude-limited sample the total number of stars expected to be observed in a (heliocentric) distance range between $D$ and $D+\mbox{d}D$ is 
\begin{equation}
\mbox{d}N(D)=\Omega\mbox{d}DD^2\rho(D)\int\limits_{M_{\min}(D)}^{M_{\max}(D)}\mbox{d}M\Phi(M)\,,
\end{equation}
where $\Omega$ denotes the solid angle covered by our sample, $\rho(D)$ the density distribution, and $\Phi(M)$ the stellar LF. The absolute magnitudes $M_{\min}(D)$ and $M_{\max}(D)$ are given by Eq.~\ref{Mmin} and \ref{Mmax}, respectively. The effective depth $D_{\mbox{\tiny eff}}$ is then given by the median value of the distribution  $\mbox{d}N(D)$ \citep{1988ApJ...329..253W}
\begin{equation}\label{Deff}
\frac{1}{2}=\frac{\int_0^{D_{\mbox{\tiny eff}}}\mbox{d}N(D)}{\int_0^{\infty}\mbox{d}N(D)}=\frac{\int_{D_{\min}}^{D_{\mbox{\tiny eff}}}\mbox{d}N(D)}{\int_{D_{\min}}^{D_{\max}}\mbox{d}N(D)}\,,
\end{equation}
where $D_{\min}$ and $D_{\max}$ are used for the numerical integration and are chosen to bracket the theoretical distance range probed by our sample, which -- neglecting interstellar dust extinction -- is readily found from the apparent magnitude limits, $m_{\min}$ and $m_{\max}$, and the absolute magnitudes limits, $M_{\min}$ and $M_{\max}$, given by $\Phi(M)$. We take them to be
\begin{equation}
D_{\min}=10^{(m_{\min}-M_{\max}+5)/5}=10^{(15-21.5+5)/5}\simeq 0.5~\mbox{pc}
\end{equation}
and
\begin{equation}
D_{\max}=10^{(m_{\max}-M_{\min}+5)/5}=10^{(20.5-(-0.5)+5)/5}\simeq 160~\mbox{kpc}\,.
\end{equation}
In Table \ref{subsamParam} we list the values of $D_{\mbox{\tiny eff}}$ for our various (sub)samples calculated using the Galactic structure parameter set 2.
With the notation of \S\ref{rhoD}, the effective galactocentric distance $\widetilde{D}_{\mbox{\tiny eff}}$ can be expressed as
\begin{eqnarray}
\widetilde{D}_{\mbox{\tiny eff}}^2&=&r_0^2+D_{\mbox{\tiny eff}}^2\cos^2b-2r_0D_{\mbox{\tiny eff}}\cos b\cos\ell\nonumber\\
&&\quad+z_0^2+2z_0D_{\mbox{\tiny eff}}\sin b+D_{\mbox{\tiny eff}}^2\sin^2b\,.
\end{eqnarray}
Its values are typically somewhat above 8 kpc. 

The last parameter we need to specify in order to estimate the Galactic tidal limit $a_{\rm T}$, is the total mass of the binary star ${\mathsf M}$. Given the magnitude limits of our sample we may write the average mass of a star in the sample as
\begin{equation}
\langle{\mathsf m}\rangle=\frac{\int\limits_{{\mathsf m}_{\min}}^{{\mathsf m}_{\max}}{\mathsf m}\xi({\mathsf m})\mbox{d}{\mathsf m}}{\int\limits_{{\mathsf m}_{\min}}^{{\mathsf m}_{\max}}\xi({\mathsf m})\mbox{d}{\mathsf m}}=\frac{\int\limits_{M_{\min}}^{M_{\max}}{\mathsf m}(M)\Phi(M)\mbox{d}M}{\int\limits_{M_{\min}}^{M_{\max}}\Phi(M)\mbox{d}M}\,,
\end{equation}
where $\xi({\mathsf m})$ denotes the mass function and ${\mathsf m}(M)$ the mass-luminosity relation which we take from \citet{1993MNRAS.262..545K}. The transformation into $r$ magnitudes was performed as in \S\ref{sec:LF}. 

Statistically, we expect the total mass of a binary star to be ${\mathsf M}=2\langle{\mathsf m}\rangle$, and we finally estimate the Galactic tidal limit according to
\begin{eqnarray}
a_{\rm T}(\widetilde{D}_{\mbox{\tiny eff}},{\mathsf M})&\simeq&\left(\frac{{\mathsf M}}{3{\mathcal M}(<\!\widetilde{D}_{\mbox{\tiny eff}})}\right)^{\frac{1}{3}}\widetilde{D}_{\mbox{\tiny eff}}\nonumber\\
&\simeq&3.1\cdot 10^{-3}\left(\frac{{\mathsf M}}{{\mathcal M}_{\odot}}\right)^{\frac{1}{3}}\left(\frac{\widetilde{D}_{\mbox{\tiny eff}}}{\mbox{pc}}\right)^{\frac{2}{3}}\mbox{pc}\,.
\end{eqnarray}
The values of $a_{\rm T}$ are around 1 pc and we list them for various (sub)samples in Table \ref{subsamParam}.

\begin{acknowledgements}
     We are grateful to P.~M.~Garnavich for providing us with his PhD thesis, which served as a starting point for our study. It is a pleasure to thank R.~Buser, B.~Krusche, T.~Lisker, N.~Sambhus, I.~Sick, G.~A.~Tammann, C.~von Arx, A.~Willand, and T.~Zingg for helpful comments and valuable discussions of the present work. We are much obliged to the referee for a prompt and insightful report. We are indebted to N.~Peduzzi and the A\&A language editor J.~Adams for proofreading our manuscript.
      
          Funding for the SDSS and SDSS-II has been provided by the Alfred P. Sloan Foundation, the Participating Institutions, the National Science Foundation, the U.S. Department of Energy, the National Aeronautics and Space Administration, the Japanese Monbukagakusho, the Max Planck Society, and the Higher Education Funding Council for England. The SDSS Web Site is http://www.sdss.org/.

    The SDSS is managed by the Astrophysical Research Consortium for the Participating Institutions. The Participating Institutions are the American Museum of Natural History, Astrophysical Institute Potsdam, University of Basel, University of Cambridge, Case Western Reserve University, University of Chicago, Drexel University, Fermilab, the Institute for Advanced Study, the Japan Participation Group, Johns Hopkins University, the Joint Institute for Nuclear Astrophysics, the Kavli Institute for Particle Astrophysics and Cosmology, the Korean Scientist Group, the Chinese Academy of Sciences (LAMOST), Los Alamos National Laboratory, the Max-Planck-Institute for Astronomy (MPIA), the Max-Planck-Institute for Astrophysics (MPA), New Mexico State University, Ohio State University, University of Pittsburgh, University of Portsmouth, Princeton University, the United States Naval Observatory, and the University of Washington.
    
    This work was supported by the Swiss National Science Foundation. 
\end{acknowledgements}

\bibliographystyle{aa}
\bibliography{13109}

\end{document}